\documentclass[graybox]{svmult}

\usepackage{type1cm}        %
\usepackage{makeidx}         %
\usepackage{graphicx}        %
\usepackage{multicol}        %
\usepackage[bottom]{footmisc}%
\usepackage{tabularx}

\usepackage{newtxtext}       %
\usepackage[varvw]{newtxmath}       %

\usepackage{hyperref}

\usepackage{svg}

\usepackage[style=numeric-comp, maxbibnames=5, maxcitenames=5,sorting=none]{biblatex}
\makeindex             %

\begin{document}

\title*{Bayesian perspectives for quantum states and application to {\em ab initio} quantum chemistry}
\titlerunning{Bayesian states for quantum chemistry}
\author{Yannic Rath\orcidID{0000-0002-4790-7422}, \\  Massimo Bortone\orcidID{0009-0005-3312-8099}, \\ George H. Booth\orcidID{0000-0003-2503-4904}
}

\institute{
Yannic Rath \at National Physical Laboratory, Teddington, TW11 0LW, United Kingdom, \email{yannic.rath@npl.co.uk}
\and Massimo Bortone \at Department of Physics, King's College London, Strand, London WC2R 2LS, United Kingdom
\and
George H. Booth \at Department of Physics, King's College London, Strand, London WC2R 2LS, United Kingdom \email{george.booth@kcl.ac.uk}
}
\maketitle

\abstract{The quantum many-electron problem is not just at the heart of condensed matter phenomena, but also essential for first-principles simulation of chemical phenomena. Strong correlation in chemical systems are prevalent and present a formidable challenge in the simulation of these systems, while predictive phenomena in this domain often also requires a demanding level of accuracy to inform chemical behavior. Efficient representations of the many-electron states of chemical systems are therefore also being inspired by machine learning principles to provide an alternative to established approaches.
  In this chapter, we review recent progress in this endeavor for quantum chemical problems represented in second quantization, and the particular challenges present in this field.
  In particular, we focus on the application of Gaussian Process States emerging from efficient representations of the many-body wavefunction with rigorous Bayesian modeling frameworks, allowing for the unification of multiple paradigms under a common umbrella.
  We show how such models (and other representations derived from machine learning) can be used as novel tools to compute \emph{ab initio} chemical properties, while in turn also informing the design of machine learning models to extract correlation patterns in classical data.
}

\vspace{1cm}

\section{The quantum chemical challenge}
\label{sec:quantum_chemistry}

Computational chemistry is a mature field with a number of distinct challenges, but of central importance is an accurate description of the ground and low-lying energetic states of the time-independent (Born-Oppenheimer) molecular Hamiltonian, given as
\begin{equation}
  H = \sum_i^N  \left( -\frac{1}{2}\nabla^2_i + v_{\text{ext}} \right) + \sum_{i<j}^N \frac{1}{|r_1-r_2|} + E_{\text{NN}} , \label{eq:firstquantH}
\end{equation}
where $N$ is the number of electrons in the system, $E_{\text{NN}}$ refers to the Coulomb repulsion between the classical point-charge nuclei, and we assume atomic units throughout.
For simplicity we have neglected relativistic effects, enforced a separability between the nuclear and electronic eigenstates (the Born-Oppenheimer approximation), and neglected any time-dependent phenomena -- all of which give rise to a number of other important challenges in computational chemistry outside the current scope. Furthermore, we can assume that $v_{\text{ext}}$ refers to the potential generated by the fixed nuclear charges, although the effect of external static fields could also be included here. Despite this list of restrictions, the solution to this Hamiltonian nevertheless encompasses a diverse array of outstanding problems in quantum chemistry, with immense commercial and industrial relevance in the development of pharmaceuticals and computationally-aided understanding of complex chemical reactions and behavior.

Increasingly, the field is aware of the limitations of existing techniques for approximating the many-electron problem above, and the prevalence of strong correlation effects in chemical systems that renders established (e.g. density functional) approaches fundamentally ill-suited~\cite{cohenChallengesDensityFunctional2012a,doi:10.1021/jacs.2c13042,D4FD00141A}. These, for example, include active sites of certain enzymes, where transition metal motifs (which often mirror those found in condensed matter systems) play a critical role in the catalytic properties of the molecule~\cite{liElectronicLandscapePcluster2019,sharmaLowenergySpectrumIron2014,kurashigeEntangledQuantumElectronic2013}. These strong correlation effects arise when the energy scale of the electron-electron repulsion for valence electrons (second term in Eq.~\ref{eq:firstquantH}) is competitive or large compared to the single-electron kinetic energy and external potential terms. Similar to strongly-correlated materials, 3$d$ valence electrons of transition metals and inorganic systems are a rich source of strong correlation effects in chemical species due to these tightly-bound local valence orbitals and their small hybridization rendering a strong electronic repulsion term. This can give rise to emergent magnetic (e.g. super-exchange) interactions between the metal atoms and a number of competing low-energy states.
However, these are far from the only examples of strong correlation in chemical systems, which are also ubiquitous at points of conical intersections (essential for non-radiative photochemical processes), bond-breaking and transition states which are crucial for a predictive approach to chemical reactions~\cite{D4FD00141A}.

\subsection{Second quantization}

A broad introduction of quantum chemical models in second quantization can be found in Ref.~\cite{szaboModernQuantumChemistry2012}, encompassing methods which traditionally project the electronic Hamiltonian into a basis set. We briefly review this framework here, stressing the aspects which affect the applicability of machine learning (ML) models in this context.

Since in chemical systems the electrons are generally localized close to nuclear centers, these basis functions are invariably (with exceptions~\cite{10.1063/5.0004792,doi:10.1137/15M1026171,PhysRevLett.119.046401,10.1063/1.4961301}) taken to be atom-centered functions, with contracted Gaussian functions the standard choice due to their ease of evaluating matrix elements between these functions with the Hamiltonian of Eq.~\ref{eq:firstquantH}. These primitive Gaussian functions are optimized and tabulated for each atom to satisfy various criteria (e.g. compactness, condition number of their overlap matrix, suitability for different properties, use with corresponding pseudopotentials, use with different levels of theory) in accessible online databases~\cite{https://doi.org/10.1002/qua.24355}. Their use, along with a number of different programs which can efficiently compute their matrix elements~\cite{10.1063/5.0200293}, simplifies both the development and application workflow in computational chemistry. The result is an easily accessible {\em second quantized} representation of the molecular Hamiltonian, as
\begin{equation}
  H = \sum_{ij} t_{ij} \hat{a}_i^\dagger \hat{a}_j + \frac{1}{2} \sum_{ijkl} v_{ijkl} \hat{a}_i^\dagger \hat{a}_j^\dagger \hat{a}_l \hat{a}_k + E_{\text{NN}} , \label{eq:secondquantH}
\end{equation}
where the tensors $t_{ij}$ and $v_{ijkl}$ define the matrix elements of the one- and two-electron terms of Eq.~\ref{eq:firstquantH} with respect to two and four basis functions respectively, and $\hat{a}^{(\dagger)}$ are fermionic annihilation (creation) operators obeying the appropriate commutation relations. Assuming a basis set of dimension $L$, the number of basis functions (in general) therefore scales as $\mathcal{O}[L^4]$. Once the basis is chosen, it is typical to exploit the flexibility to linearly transform the degrees of freedom to change the representation of the single-particle states while preserving their span. This can be done to ensure that the single-particle functions are orthonormal (and therefore that the many-body basis functions comprised of their tensor product are orthonormal), as well as requiring that the functions represent eigenstates of some effective one-body Hamiltonian (such as found from Hartree--Fock or density functional theory). This often allows for a large degree of sparsity in the many-body amplitudes if this one-body Hamiltonian represents a qualitatively correct starting point~\cite{BYTAUTAS200964}. We will generically call these degrees of freedom as `orbitals', noting that there is substantial freedom in their precise choice, and could be localized or delocalized single-particle states. The freedom to choose this representation will be discussed more in Sec.~\ref{sec:quantchem_challenge}.

While quantum chemical methods almost ubiquitously start from this second quantized Hamiltonian, traditional quantum Monte Carlo approaches for {\em ab initio} Hamiltonians have favored the first quantized representation of Eq.~\ref{eq:firstquantH}. This includes recent advances in machine-learning inspired quantum many-electron states in first quantization, such as FermiNet~\cite{Pfau2019}, PauliNet~\cite{Hermann2019} and others~\cite{Fu_2024}. These approaches typically rely on an antisymmetrized product of single-electron orbitals (or short linear combinations thereof), where each orbital also depends on all other electron coordinates in a permutation-equivariant fashion. This dependence on all electron coordinates necessitates a compact yet flexible and improvable functional form for these orbitals, which therefore motivates the parameterization via machine learning models, such as neural networks. The approach overall is akin to previous work on `backflow' wavefunctions for the description of correlated physics in Fermionic systems, which more traditionally used fixed parameterizations via a low-body expansion~\cite{tocchioRoleBackflowCorrelations2008,tocchioBackflowCorrelationsHubbard2011}.

The success of first-quantized approach to {\em ab initio} quantum Monte Carlo is primarily due to the fact that since Eq.~\ref{eq:firstquantH} only depends on a sum of two electron coordinates, the evaluation of the local energy for a particular electronic configuration can be performed efficiently -- often as low as $\mathcal{O}[N^3]$ depending on the complexity of the wavefunction model~\cite{Becca2017}. In contrast, the complexity of Eq.~\ref{eq:secondquantH} ensures that the brute-force evaluation of the local energy for a given second quantized electron configuration (a string of occupations of each degree of freedom in the basis) scales at least as $\mathcal{O}[L^4]$ and often worse, depending on the complexity of the model evaluation, and noting that $L$ scales linearly with $N$ (and is at least as large). Coupled to this, the imposition of a basis set of one-body functions in which the many-body Hilbert space is expanded necessarily leads to a truncation in the span of physical states which can be described. This expansion of the basis generally first describes the low-energy and strong quantum fluctuations of the system, while larger basis sets are required for quantitative accuracy derived from the high-energy, two-body physics of electron scattering processes at the shortest length scales between electronic coordinates. These latter processes require large basis sets to describe these wavefunction features, with the wavefunctions exhibiting a universal sharp derivative discontinuity (`cusp') around all inter-electronic coaleascence points, independent of the details of the chemical environment~\cite{doi:10.1021/cr200204r}. This choice of basis necessarily imposes a low-energy truncated space in which the wavefunction is expanded, leading to a slowly decaying convergence in the correlated energy expectation value, which decays as $\mathcal{O}[L^{-1}]$, regardless of the choice of specific function for the basis set expansion.

Despite this drawback, there are also some advantages to a second quantized representation which we believe make their continued development in the framework of machine-learning inspired quantum state representations (and stochastic optimization) worthwhile. Firstly, permutational invariance of the wavefunction amplitudes with respect to the electrons is automatically enforced on the level of the operators, along with the requirements of overall antisymmetry with respect to pairwise electron permutations. This non-local constraint can be considered as equivalent to a Jordan-Wigner transformation of the fermions to spin degrees of freedom, though other mappings to spin models can also be considered~\cite{Choo2019a}. This means that explicit constraints on permutational invariance and antisymmetry no longer need to be considered in the state definition, noting however that the choice of ordering of the fermionic operators still affects the exact probability amplitudes -- a point we return to in Sec.~\ref{sec:quantchem_challenge}.

This invariance means that the exact state within the basis set can be found via a constraint-free variational optimization of the probability amplitudes via the Raleigh-Ritz energy functional, resulting in the exact diagonalization or `full configuration interaction' (FCI) approach. This provides a `ground-truth' set of amplitudes, representing the state as an $L$-indexed tensor of probability amplitudes,
\begin{equation}
  |\Psi\rangle = \sum_{n_1, n_2, n_3, \ldots, n_L} \Psi_{n_1, n_2, n_3, \ldots, n_L} |\vec{n} \rangle, \label{eq:FCI}
\end{equation}
where we associate an amplitude $ \Psi_{n_1, n_2, n_3, \ldots, n_L}$ to each computational basis state $|\vec{n}\rangle = |n_1, n_2, n_3, \ldots, n_L \rangle$.
Here, we define the computational basis via the possible configurations of electrons in the molecular orbital space.
The Fermionic character dictates that each orbital can only be occupied by up to two electrons of opposing spins, giving four possible local occupancies defining the Fock space of each orbital, $n_i \in \{\cdot, \uparrow, \downarrow, \downarrow \uparrow \}$. Therefore, as long as the probability amplitudes of an arbitrary model for a given electronic configuration can be efficiently evaluated, second quantization lends itself to a broad flexibility in the models considered, including matrix product states (MPS)~\cite{10.1063/1.4955108}, correlator product states (CPS)~\cite{Clark_2018}, as well as a number of machine-learning inspired forms~\cite{Carleo2017}.

Second-quantized representations of quantum states also allow for a particular ease in a multi-resolution methodology to enable applicability to larger systems. This could, for example, allow the use of accurate yet computationally demanding levels of theory in smaller Hilbert spaces in order to capture important correlated physics of a system. This partitioning of a larger Hilbert space into smaller pieces can be achieved in either an energy or real-space domain, giving rise to `active space' methods in the former~\cite{ROOS2005725}, or quantum embedding methods for the latter, such as dynamical mean-field theory or density matrix embedding~\cite{doi:10.1021/acs.accounts.6b00356, doi:10.1021/acs.jctc.0c01258,PhysRevX.12.011046}. This ease in which arbitrary second-quantized `solvers' can fit into a wider multi-resolution workflow is particularly appealing for their development, such that the Hilbert space corresponding to only the most challenging parts of the quantum many-body problem can be treated in isolation.
This second-quantized multi-resolution approach also extends to a straightforward and flexible treatment of core electrons e.g. via effective core potentials~\cite{10.1063/1.448975}, or approaches to ameliorate the challenge of reaching the basis set convergence, such as including explicit correlation to account for the sharp basis set features via techniques such as transcorrelation, canonical transformations or `F12' methods, largely mitigating the basis set approximation in second quantized techniques~\cite{doi:10.1021/cr200204r,10.1063/1.3688225,doi:10.1098/rspa.1969.0120,10.1063/1.4976974,10.1063/1.4959245,10.1063/1.4762445}.

Finally, we address the fact that the local energy evaluation appears to be so much more computationally costly in a second quantized representation compared to first quantization. We note that this additional scaling must be somewhat artificial, since the Hamiltonian in second quantization is still ultimately a sum of pairwise operators. Indeed, we find that in a local representation of the degrees of freedom, the locality of each electron considered ensures that the two-electron operator in Eq.~\ref{eq:secondquantH} asymptotically only has $\mathcal{O}[N^2]$ terms. This can be exploited to ensure that the formal scaling of the local energy for a given electronic configuration is the same for the two representations~\cite{weiReducedScalingHilbert2018}.

In the rest of this chapter, we will consider the use of Bayesian-inspired wavefunction models in second quantization for chemical application. We first give an overview of Bayesian-inspired ansatzes in the following two sections before returning to their application in the context of quantum chemistry in section~\ref{sec:quantchem_challenge}.

\section{Bayesian machine-learning inspired quantum states}
\label{sec:Bayesian_wavefunction}

\subsection{Modeling many-body wavefunctions}
\begin{figure}
  \centering
  \def\svgwidth{\textwidth}
\begingroup%
  \makeatletter%
  \providecommand\color[2][]{%
    \errmessage{(Inkscape) Color is used for the text in Inkscape, but the package 'color.sty' is not loaded}%
    \renewcommand\color[2][]{}%
  }%
  \providecommand\transparent[1]{%
    \errmessage{(Inkscape) Transparency is used (non-zero) for the text in Inkscape, but the package 'transparent.sty' is not loaded}%
    \renewcommand\transparent[1]{}%
  }%
  \providecommand\rotatebox[2]{#2}%
  \newcommand*\fsize{\dimexpr\f@size pt\relax}%
  \newcommand*\lineheight[1]{\fontsize{\fsize}{#1\fsize}\selectfont}%
  \ifx\svgwidth\undefined%
    \setlength{\unitlength}{453.53803019bp}%
    \ifx\svgscale\undefined%
      \relax%
    \else%
      \setlength{\unitlength}{\unitlength * \real{\svgscale}}%
    \fi%
  \else%
    \setlength{\unitlength}{\svgwidth}%
  \fi%
  \global\let\svgwidth\undefined%
  \global\let\svgscale\undefined%
  \makeatother%
  \begin{picture}(1,0.36607956)%
    \lineheight{1}%
    \setlength\tabcolsep{0pt}%
    \put(0,0){\includegraphics[width=\unitlength,page=1]{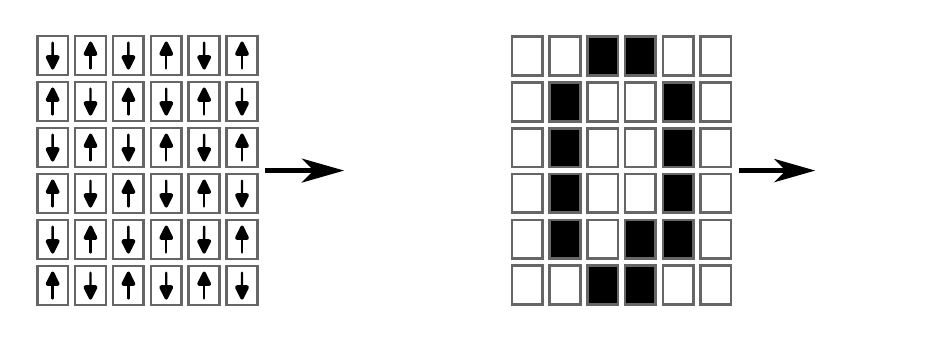}}%
    \put(0.37361604,0.17250992){\color[rgb]{0,0,0}\makebox(0,0)[lt]{\lineheight{1.25}\smash{\begin{tabular}[t]{l}{\Large$\Psi_\mathrm{AFM}$}\end{tabular}}}}%
    \put(0.8722533,0.17250992){\color[rgb]{0,0,0}\makebox(0,0)[lt]{\lineheight{1.25}\smash{\begin{tabular}[t]{l}{\Large $p_0$}\end{tabular}}}}%
  \end{picture}%
\endgroup%
   \caption{Analogy between the structure emerging within physical many-body wavefunctions and in pixel arrays representing a handwritten digit~\cite{rath_thesis}. An antiferromagnetic state will associate a large wavefunction amplitude to the configuration where spins on neighbouring sites are aligned anti-parallel (left), similarly only specific pixel configurations have a high probability of representing the handwritten digit "0" (right).}
  \label{fig:wavefunction_ML_analogy}
\end{figure}

As stated in Eq.~\ref{eq:FCI}, a valid second quantized wavefunction model is entirely specified by a set of complex-valued amplitudes, $\Psi_{n_1, n_2, n_3, \ldots, n_L}$, for each electronic configuration in the exponentially large Hilbert space. Each computational basis state representing this electronic configuration can be specified by a bitstring of $2L$ bits, which encode the specific occupations, $n_i$, of each of the Fermionic orbitals.
Electronic wavefunctions are not entirely random states in the Hilbert space but exhibit a large degree of physical structure, much like arrays of pixels become structured as these represent objects in a digital image (see Fig.~\ref{fig:wavefunction_ML_analogy}).

One of the most central questions within the field of computational quantum chemistry is what exactly this emergent structure is within many-electron states of interest, and how this can be exploited to define a compact function approximator to the wavefunction $\Psi(\vec{n}) = \Psi_{n_1, n_2, n_3, \ldots, n_L}$ that enables practically efficient computations. This allows us to circumvent the requirement to store all of these exponentially-many amplitudes explicitly and independently, as done in FCI.
It is clear that the structure of the wavefunction amplitudes depends on the system under study, where different system interactions can give rise to very different characteristics.
Standard ansatzes for the quantum state derived from physical intuition thus typically only capture a specific type of physical behaviour, and these will not cover all emergent structure in these amplitudes. However, universal guiding principles of locality and a low-rank nature of the correlated features can be exploited to devise starting forms, from which systematic expansions beyond these can be expected to be rapidly convergent.

The starting point in chemical systems is generally a mean-field treatment, giving the state as a single Slater determinant. Quantum fluctuations about this starting point are however required for an accurate description of the neglected electron correlations in this mean-field state. %
An important characteristic of a model capturing the correlation is product separability.
This ensures that the model can correctly represent the state over non-interacting subsystems as a product state over the state of each subsystem. This is a property of the exact eigenstates of the Hamiltonian, which also factorizes as a product over non-interacting subsystems.
Incorporating a product structure over the system constituents is therefore a crucial ingredient to ensure the `size-extensivity' of the ansatz, such that its total energy scales extensively with system size, while the energy {\em density} is intensive~\cite{bartlettManyBodyPerturbationTheory1981,https://doi.org/10.1002/wcms.1120}~\footnote{Strictly, product separability implies size {\em consistency}, ensuring that the sum of the energies of non-interacting subsystems is the same as the energies of those isolated subsystems, rather than the energy extensivity requirement. However, these points are often related~\cite{https://doi.org/10.1002/wcms.1120}.}.

A prominent wavefunction ansatz which explicitly builds upon this multiplicative concept is the family of `Correlator Product States' (CPS)~\cite{Changlani2009}, also known as `Entangled Plaquette States'~\cite{mezzacapoGroundstatePropertiesQuantum2009}.
This family of states defines a many-body ansatz for the wavefunction as a product over $N_p$ (possibly overlapping) sets of orbitals or plaquettes, where a full parameterization of independent amplitudes over the configurations within each plaquette is optimized.
This ansatz for the overall probability amplitudes can then be written as a product of the amplitudes of each plaquette, as
\begin{equation}
  \Psi_\mathrm{CPS}(\vec{n}) = \prod_{i=1}^{N_p} \psi^{(i)}_{n_{p_1(i)}, \ldots, n_{p_P(i)}},
\end{equation}
where each amplitude tensor $\psi^{(i)}_{n_{p_1(i)}, \ldots, n_{p_P(i)}}$ encodes the amplitudes across the $i$-th plaquette of orbitals with indices $\{p_1(i) \ldots p_P(i)\}$, noting that different plaquettes could have different numbers of orbitals in its definition.
The standard CPS functional form as introduced above explicitly relies on parameterizing the full Hilbert space for each plaquette.
By tiling these finite-sized plaquettes across the full system, a compact parametrization can be achieved by factorizing the wavefunction amplitudes over these plaquette contributions (which are generally taken to be comprised of spatially local degrees of freedom). This form makes it possible to model specific correlation properties efficiently. This concept of correlating plaquettes will be the starting point for the machine-learning inspired Gaussian Process State model.

\subsection{The Gaussian Process State as a Bayesian model of wavefunctions}

While a CPS ansatz can be systematically improved to describe any quantum state of interest by increasing the size of the plaquettes, the representation quickly becomes intractable as its complexity grows exponentially in the size of the plaquettes.
Furthermore, there is no general recipe to design plaquettes to achieve the best possible approximation for a given problem.
Here, the application of data-driven techniques provides an appealing alternative.
Rather than constraining the model to a certain type of physical behaviour, such approaches hold the promise to extract the most relevant correlation properties of many-body quantum states automatically.
The application of data-driven approaches builds upon the representational power of machine learning models able to capture the structure of high-dimensional input-output relationships, between a vectorial input configuration $\vec{n}$ and an output function $f(\vec{n})$.
The modelling of many-body wavefunctions is therefore closely related to tasks of supervised learning, where an efficient description of the input-output relationship is inferred based on few observations.

Perhaps one of the most intuitive approaches for the modeling of high-dimensional data is a simple linear function in a suitable (often higher-dimensional) space of features.
Such a function approximator may be defined as
\begin{equation}
  f_\mathrm{lin}(\vec{n}) = \sum_{i=1}^{N_\mathrm{features}} w_i \, \chi_i(\vec{n}),
  \label{eq:linear_model}
\end{equation}
where a linear combination of the $N_\mathrm{features}$ feature transformations, $\chi_i(\vec{n})$, map each input $\vec{n}$ to a scalar quantity, generally in a highly non-linear fashion.
The weights $w_i$ associated with the features may be considered the parameters of the model, which can be obtained by fitting this model to data.

\begin{figure}[]
  \centering
  \includegraphics[width=0.8\textwidth]{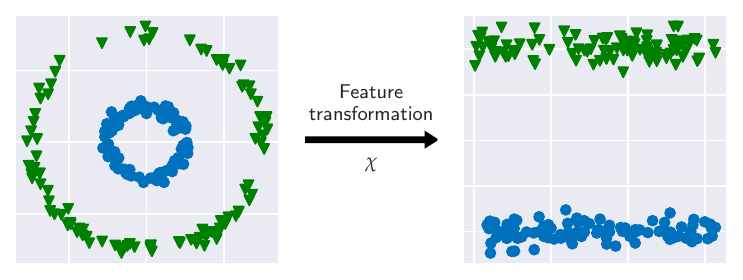}
  \caption{A non-linear transformation of the original input data to a feature space allows for linear separability of the input data~\cite{rath_thesis}. The plots show two sets of data points associated with two different classes (blue and green). The left panel shows the data in the original space of Cartesian coordinates, the right plot shows the same data in a feature space given by polar coordinates in which the two classes are now linearly separable.}
  \label{fig:feature_space}
\end{figure}

A classic example of the importance of the feature transformation is exemplified for the classification of the data shown in Fig.~\ref{fig:feature_space}.
In the figure, two classes of data points (blue and green) are positioned in the space of Cartesian coordinates, which cannot be separated by a linear function in this original input space.
Due to the radial distribution of the data points however, it is possible to linearly separate the two classes in the space of polar coordinates.
Changing from Cartesian to polar coordinates is an example of a specific non-linear feature transformation of the inputs into a two-dimensional feature space. Once transformed, it can be seen that a linear model is sufficient to correctly classify the data points.

While in the example of transforming the inputs into polar coordinates the feature space is only two-dimensional, in general the feature space can be of much higher dimensionality than the input space.
This makes it possible to define a general model, which can represent any function to arbitrary accuracy as the number of features is increased.
However, the main question of practical importance is how to design these features such that a compact model can be found from limited data that also generalizes well.
We therefore need to define the set of feature transformations, and an appropriate protocol to optimize the weights of the subsequent model for a target input-output relationship (such as an electronic wavefunction).

A common choice for the practical application of linear models in a systematically-improvable way is the utilization of an exponentially large set of features which are designed to be `complete', such that they can represent a true universal approximator.
While this may appear to render the evaluation of such a model according to Eq.~\eqref{eq:linear_model} impractical, only scalar products between vectors of features associated with different inputs need to be evaluated when fitting such a model to a data set~\cite{Rasmussen2006}.
This is the core component of \emph{kernel models}, where the feature transformation is only defined implicitly through a positive semi-definite kernel function $k(\vec{n}, \vec{n}')$, representing the scalar product between feature vectors associated with two inputs $\vec{n}$ and $\vec{n}'$, with $k(\vec{n}, \vec{n}')=\sum_{i}^{N_\text{features}} \chi_i (\vec{n}) \cdot \chi_i(\vec{n}')$.
This `kernel trick' gives rise to a model of the form
\begin{equation}
  f_\mathrm{kernel}(\vec{n}) = \sum_{\vec{n}'} \tilde{w}_{\vec{n}'} \, k(\vec{n}, \vec{n}'), \label{eq:kernelmodel}
\end{equation}
where a data set of configurations $\{\vec{n}'\}$ explicitly enters the model definition, together with the kernel function weighted by coefficients $\tilde{w}_{\vec{n}'}$.

Due to their essentially unlimited representational power, yet still allowing for a high degree of interpretability, kernel models have also proven to be well suited to extract the structure of many-body wavefunctions~\cite{Glielmo2020,Rath2020,https://doi.org/10.48550/arxiv.2303.08902}.
As scalable wavefunctions are required to factorize as products over all sub-regions of a system, a practical useful ansatz constructed from a kernel model should also satisfy this product structure. However, it can be seen that kernel models are instead additively separable over configurational features (which have yet to be defined).
Following the construction of other common wavefunction models, such as Jastrow ansatzes~\cite{jastrowManyBodyProblemStrong1955}, a product structure for the wavefunction amplitudes can be therefore obtained by an exponentiation of the form given in Eq.~\ref{eq:kernelmodel}.
This defines the `Gaussian Process State' (GPS) ansatz~\cite{Glielmo2020,Rath2020}, owing its name to the Gaussian process regression framework -- a supervised learning paradigm building upon rigorous statistical principles, in which a kernel model is extracted from a given data set in a probabilistic sense~\cite{Rasmussen2006}.

The GPS therefore defines a general-purpose wavefunction ansatz as an exponentiated kernel model.
It associates a wavefunction amplitude to a computational basis state according to the form
\begin{equation}
  \Psi_\mathrm{GPS}(\vec{n}) = e^{\sum_{i=1}^{M} w_i \, k(\vec{n}, \vec{n}'_i)},
\end{equation}
where the index $i$ labels $M$ different electronic configurations $\vec{n}'_i$, which explicitly enter the model definition as a data set of `support configurations'.
Together with these support configurations and the kernel function, the model is fully specified through the weights, $w_i$, which can be interpreted as continuous (in general complex-valued) variational parameters of the model.
The following sections outline how physically intuitive and complete kernel functions can be constructed, and how particularly relevant features can be identified from given wavefunction data to give insight into the emergent properties of the many-electron state based on Bayesian regression frameworks.

\subsection{Kernel functions for Gaussian Process States} \label{sec:kernel}

Leaving the task of selecting an appropriate set of support configurations aside, the expressibility of the GPS is governed by the choice of the kernel function, which can be loosely interpreted as the covariance of the prior probability distribution between points in the input space from a Bayesian perspective.
The kernel function simply takes two configurations $\vec{n}$ and $\vec{n}'$ from the Hilbert space as its input and maps these to a scalar quantity $k(\vec{n}, \vec{n}')$.
Following the discussion above, it can also directly be identified as a scalar product in the space of modelled correlation features (in which case it should be a symmetric positive semi-definite function).
The kernel function therefore implicitly defines the features of the correlated physics that are modelled (and a weighted importance attributed to them).
By modelling the log wavefunction amplitudes as a linear model, the resulting GPS can be interpreted as a product over the weighted correlation features, in a similar spirit to the construction of a CPS as a product over plaquettes.
However in contrast to the CPS, by choosing a suitable kernel function, the GPS can capture arbitrary correlations through these features across the system, without the requirement of specifying a restricted set of plaquettes to explicitly define the rank and range of these correlations.

The construction of an `exponential kernel' in Refs.~\cite{Glielmo2020, Rath2020} builds this fully flexible kernel (analogously to a discrete version of a squared-exponential kernel) by systematically weighting features defined by the configuration within a given plaquette of orbitals/sites that are common between the two configurations $\vec{n}$ and $\vec{n}'$. It is possible then to analytically resum the features over {\em all} possible plaquettes of the system, including all rank, range and topology of these plaquettes, avoiding the need to define these plaquettes explicitly.
This exponential kernel is defined as
\begin{equation}
  k(\vec{n}, \vec{n}') = e^{-h(\vec{n}, \vec{n}')}.
  \label{eq:exponential_kernel}
\end{equation}
The function $h(\vec{n}, \vec{n}')$ is a weighted distance metric between the two configurations $\vec{n}$ and $\vec{n}'$, defined as a weighted Hamming distance~\cite{hammingErrorDetectingError1950}, $h(\vec{n}, \vec{n}') = \sum_{i=1}^{L} \frac{1 - \delta_{n_i, n'_i}}{f(i)}$. A central building block for this kernel is the comparison of local occupancies of computational basis states $\vec{n}$ and $\vec{n}'$ via the Kronecker delta, $\delta_{n_i, n'_i}$, which evaluates to one if the local occupancy at site $i$ is equal in the two configurations, and to zero otherwise.
The weighting function $f(i)$ can be specified by additional hyperparameters of the model and allows for control to weight different features corresponding to physical correlations which we want to be attribute more importance to in the subsequent fitting. A slowly-decaying $f(i)$ will put more emphasis on fitting high-rank features which correlate the occupations over larger numbers of physical sites, while if the function decays rapidly with the distance from a reference site, then it promotes the fitting of short-range correlated features.

\begin{figure}
  \centering
  \def\svgwidth{\textwidth}
\begingroup%
  \makeatletter%
  \providecommand\color[2][]{%
    \errmessage{(Inkscape) Color is used for the text in Inkscape, but the package 'color.sty' is not loaded}%
    \renewcommand\color[2][]{}%
  }%
  \providecommand\transparent[1]{%
    \errmessage{(Inkscape) Transparency is used (non-zero) for the text in Inkscape, but the package 'transparent.sty' is not loaded}%
    \renewcommand\transparent[1]{}%
  }%
  \providecommand\rotatebox[2]{#2}%
  \newcommand*\fsize{\dimexpr\f@size pt\relax}%
  \newcommand*\lineheight[1]{\fontsize{\fsize}{#1\fsize}\selectfont}%
  \ifx\svgwidth\undefined%
    \setlength{\unitlength}{453.53803019bp}%
    \ifx\svgscale\undefined%
      \relax%
    \else%
      \setlength{\unitlength}{\unitlength * \real{\svgscale}}%
    \fi%
  \else%
    \setlength{\unitlength}{\svgwidth}%
  \fi%
  \global\let\svgwidth\undefined%
  \global\let\svgscale\undefined%
  \makeatother%
  \begin{picture}(1,0.3200037)%
    \lineheight{1}%
    \setlength\tabcolsep{0pt}%
    \put(0,0){\includegraphics[width=\unitlength,page=1]{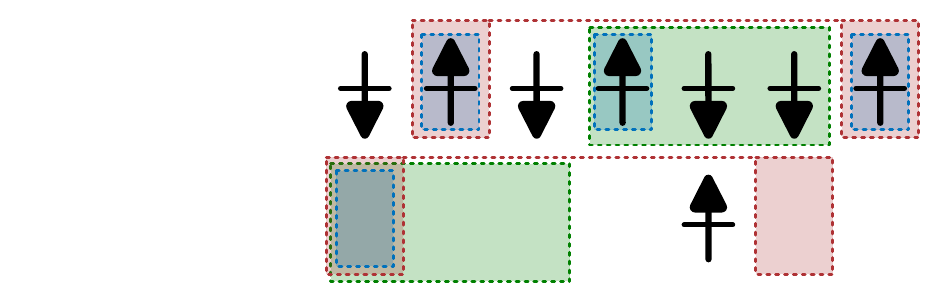}}%
    \put(0.07447265,0.22028127){\makebox(0,0)[lt]{\lineheight{1.25}\smash{\begin{tabular}[t]{l}Test configuration $\mathbf{n}$\end{tabular}}}}%
    \put(0.02470666,0.07586698){\makebox(0,0)[lt]{\lineheight{1.25}\smash{\begin{tabular}[t]{l}Support configuration $\mathbf{n}'$\end{tabular}}}}%
    \put(0,0){\includegraphics[width=\unitlength,page=2]{kernel_design.pdf}}%
  \end{picture}%
\endgroup%
   \caption{Matching of sub-configuration patterns over different plaquettes between two configurations in the exponential kernel~\cite{rath_thesis}. The final kernel value can be implicitly associated with a weighted sum of common features over these sub-configurations within all possible plaquettes. The coloured plaquettes indicate such equal sub-configurations for the two example spin configurations $\mathbf{n}$ and $\mathbf{n}'$.}
  \label{fig:kernel_design}
\end{figure}

The fact that this kernel compactly resums over the (exponential) set of all possible correlated features over all possible plaquettes is best seen by considering its Taylor expansion, giving
\begin{equation}
  k(\vec{n}, \vec{n}') = \left( 1 - \sum_{i=1}^{L-1} \frac{1 - \delta_{n_i, n'_i}}{f(i)} + \sum_{i,j=1}^{L-1} \frac{(1 - \delta_{n_i, n'_i}) (1 - \delta_{n_j, n'_j})}{f(i) \cdot f(j)} - \ldots \right).
\end{equation}
The first term thus solely extracts single-site correlation features (i.e., Gutzwiller-like features), the second term gives two-site correlation features between pairs of sites (i.e., two-site Jastrow-like features), and higher order terms give increasingly complex many-body correlation features.
The extent by which higher-order terms contribute to the overall kernel function is controlled by the weighting $f(i)$, which can be chosen to amplify or suppress the correlation features depending on the order and range~\cite{Rath2020}.
The physical rationalization of this additive kernel as a sum over features corresponding to shared plaquettes of the configurations is visualized in Fig.~\ref{fig:kernel_design}, showing how the kernel value is obtained by finding matching patterns over plaquettes for two configurations of a spin-$\frac{1}{2}$ chain~\cite{NIPS2011_4c5bde74}.

\subsection{Bayesian optimization of Gaussian Process States}

Having defined a suitable kernel function and hence an implicit space of features, we now consider how this can be used to extract a compressed representation of a target many-body state.
Firstly, we consider a supervised learning algorithm: having access to a set of amplitudes of a given quantum state, we aim to extract a sparse representation of this state as a GPS model. We consider the schemes discussed in Refs.~\cite{Rath2020, boothQuantumGaussianProcess2021}, where the training data is chosen to be samples of the ground state (FCI) wavefunction amplitudes for small systems.

The GPS defines the log amplitudes of the wavefunction amplitudes as a kernel model
\begin{equation}
  \phi_\mathrm{GPS}(\vec{n}) = \log \left( \Psi_\mathrm{GPS}(\vec{n}) \right) = \sum_{i=1}^M w_i \, k(\vec{n}, \vec{n}_i'). \label{eq:logGPSmodel}
\end{equation}
Assuming a set of $N_\mathrm{trn}$ data points (i.e. many-body configurations and their associated amplitudes), we can rely on Bayesian regression approaches to optimize the model~\cite{Tipping2003a, Rasmussen2006}. This constructs a model in a statistical sense as the `most probable' model to describe the log-wavefunction amplitudes, given the data.
To this end, we first introduce a \emph{prior} probability distribution, $p_\mathrm{prior}(\vec{w})$, over the vector of model weights $\vec{w}$, encoding our belief about the distribution of the weights in the absence of any data.
The specific data set is then taken into account through the \emph{likelihood} of observing log-wavefunction amplitudes for the data points, given the weights $\vec{w}$.
We denote this as $p_\mathrm{lik}(\vec{\phi} | \{\vec{n}\}, \vec{w})$, where $\vec{\phi}$ corresponds to the vector of observed log-amplitudes for a set of basis states $\{\vec{n}\}$.
Determining the weights can then be achieved by application of Bayes theorem, giving a posterior probability distribution for the weights as the normalised product of prior and likelihood, as
\begin{equation}
  p_\mathrm{post} = \frac{p_\mathrm{prior}(\vec{w}) \times p_\mathrm{lik}(\vec{\phi} | \{\vec{n}\}, \vec{w})}{\int \, d \vec{w} p_\mathrm{prior}(\vec{w}) \times p_\mathrm{lik}(\vec{\phi} | \{\vec{n}\}, \vec{w})}.
  \label{eq:Bayes}
\end{equation}

This Bayesian approach relies on the definition of suitable prior and likelihood distributions.
As discussed in Ref.~\cite{Rath2020}, we follow the approach of Gaussian process regression, where we assume multi-variate Gaussian distributions for likelihood and prior.
The likelihood distribution for the log-wavefunction amplitudes is assumed to follow independent Gaussian distributions around the GPS model predictions for a given set of weights, whose prior probability distribution is Gaussian with zero mean.
As the posterior distribution for the expected weights emerges from the product of two Gaussian distributions, the resulting posterior is also a normal distribution, and can be found in closed-form.
The mean of this distribution is denoted as a vector $\vec{\mu}_\mathrm{mp}$ as the most probable weights over all possible models, given by
\begin{equation}
  \vec{\mu}_\mathrm{mp} = \vec{\Sigma}\vec{K}^T \vec{B} \vec{\phi}.
  \label{eq:mean_weights}
\end{equation}
The quantity $\vec{\Sigma}$ denotes the covariance matrix of the posterior distribution and is given by
\begin{equation}
  \vec{\Sigma} = (\vec{K}^T \vec{B} \vec{K} + \vec{A})^{-1}.
  \label{eq:covariance_weights}
\end{equation}
This expression utilizes the shorthand notation $\vec{K}$ to denote the $N_\mathrm{trn} \times M$ matrix with rows corresponding to the transposed kernel values for all training inputs, and $\vec{A}$ is a matrix parametrizing the inverse of the weight prior covariance matrix.
Similarly, $\vec{B}$ is a $N_\text{trn} \times N_\text{trn}$ diagonal matrix comprising of the inverse likelihood variances, $(\sigma^2(\vec{n}_i))^{-1}$, capturing an assumed noise in the data points.
We can appropriately regularize the fit to the data amplitudes in the log space, via a choice of the amplitude-dependent `noise' variances in the likelihood function~\cite{Glielmo2020, Rath2020}, as
\begin{equation}
  \sigma^2(\mathbf{n}) = \log \left( \frac{\tilde{\sigma}^2} {|e^{\phi(\vec{n})}|^2} + 1 \right),
  \label{eq:log_space_likelihood_variance}
\end{equation}
where $\phi(\vec{n})$ is the log-amplitude of data configuration $\vec{n}$, and $\tilde{\sigma}^2$ is a hyperparameter characterizing a target accuracy for the GPS wavefunction amplitudes to fit the given data.
Adopting the most-probable weights $\vec{\mu}_\mathrm{mp}$ therefore defines a final GPS model from the data set.

\subsection{Choice of support configurations}

For a practical application of the supervised learning scheme, we also require the selection of support configurations, $\{\vec{n}'\}$, in the model definition.
This may either be the same set of training configurations from the data set used to define the weights in the Bayesian framework above, as is common in kernel models~\cite{https://doi.org/10.48550/arxiv.2303.08902}, or indeed a different set which aims to achieve a more compact model definition which we describe here.

We can once again appeal to Bayesian modeling principles to identify the most relevant configurations to choose in a data set, resulting in an optimally compact GPS model (i.e. one which minimizes the number of support configurations dubbed the `support dimension', $M$, in Eq.~\ref{eq:logGPSmodel}). This involves application of the \emph{Relevance Vector Machine} (RVM)~\cite{Tipping2000, Tipping2003} as a tool to extract a particularly compact GPS model with small support dimension~\cite{Glielmo2020, Rath2020}.
The denominator in Eq.~\ref{eq:Bayes} is denoted the marginal likelihood, and quantifies how well the quality of the fit is balanced with the complexity of the model.
This measure of trade-off between model sparsity and accuracy can be exploited to select the most suitable set of support configurations in the RVM algorithm by a maximization of the quantity with respect to a parametrized prior covariance~\cite{Tipping2000, Tipping2003}.
This way, the RVM identifies the most relevant support configurations from a set of candidates, defining a sparse set used for the definition of the wavefunction ansatz~\cite{Rath2020}.

\begin{figure}
  \centering
  \includegraphics[width=\textwidth]{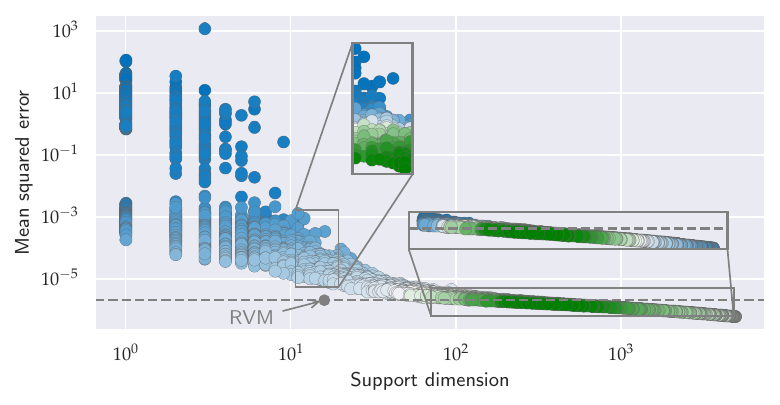}
  \caption{Mean squared error in the GPS compared to the exact state of a Hubbard model ground state, as the number of randomly selected data (`support') configurations defining the fit is increased. The GPS is fit with a symmetrized exponential kernel with an inverse-distance weighting of correlation features.
  The coloring of the scatter points denotes the marginal likelihood of the model, where larger values are represented by green, smaller values by blue, and the scale is renormalized in the two insets. The dark grey point indicates the result obtained with the application of the relevance vector machine to select the most relevant and compact set of support configurations.  The specific Hubbard hamiltonian is a half-filled one-dimensional anti-periodic chain of $L=8$ sites with a ratio of interaction to nearest-neighbor hopping of $U/t=8$. Figure modified from Ref.~\cite{Rath2020}.}
  \label{fig:support_set_selection}
\end{figure}

Figure~\ref{fig:support_set_selection} highlights the success of the RVM to select a compact set of particularly relevant support configurations for the model.
For a fit to the data from a one-dimensional Fermi-Hubbard chain~\cite{Rath2020} obtained from FCI, the figure gives the mean squared error of the GPS amplitudes compared to the exact target state for different support sets.
The error achieved in the model based on the automatic selection of support configurations by the RVM for a chosen set of hyperparameters (grey scatter point), is significantly more accurate for its small size than for different random sets of support configurations (colored scatter points).

The selection of a compact model via the RVM along with Bayesian optimization of the weights, allows for an efficient overall supervised learning protocol to map any training information over a (potentially incomplete) set of probability amplitudes of configurations into a GPS form which takes values over all configurations in the Hilbert space. The inductive bias of the model can be controlled by the hyperparameters of the model, but its motivation in terms of correlation features allows good generalization properties of the model across all configurations. Application of this supervised learning of the GPS to generalization problems such as quantum state tomography~\cite{Torlai2018a} or measurement protocols in quantum devices~\cite{doi:10.1021/acs.nanolett.4c04889,Neugebauer2020} is therefore an intriguing possibility.

\subsection{From classical support configurations to quantum product states} \label{sec:qGPS}

While the RVM can significantly sparsify the set of support configurations, an alternative method has emerged to improve the GPS model, which entirely removes the need to select discrete support configurations. Instead of defining the kernel as a distance metric between `classical' configurations of the particles amongst the degrees of freedom, Ref.~\cite{boothQuantumGaussianProcess2021} introduces a parametrization of these support configurations as general product states (i.e. superpositions of the local occupancies of each site).
By co-optimizing the continuous parameters defining these product states along with the weights when fitting the model to data, the RVM compression scheme is no longer required to select the configurations since the support states can continuously deform from one state to another.
The support dimension of the model is therefore kept fixed and treated as an additional hyperparameter of the model, which also allows for a direct control of the model complexity.

To derive this form, we can expand the exponential kernel defined in Sec.~\ref{sec:kernel} into a product over sites, allowing the log wavefunction amplitudes to be written as
\begin{equation}
  \phi(\vec{n}) = \sum_{i=1}^M w_i \left ( \prod_{j=1}^L k^{(j)}(n_j, n'^{(i)}_j) \right),
\end{equation}
where $n'^{(i)}_j$ denotes the occupancy of the $i$-th support configuration at site $j$, and $k^{(j)}$ defines a local kernel function comparing these local occupancies to the test configuration $\vec{n}$. Using Eq.~\ref{eq:exponential_kernel}, these local kernel values are given by
\begin{equation}
  k^{(j)}(n_j, n'_j) =    e^{\frac{-(1-\delta_{n_j, n'_j})}{f(j)}}.
\end{equation}
However, we can now let the local kernel for a site $j$ and a support configuration $\vec{n}'$ take a more flexible form, whereby the binary classification comparing the occupancies at a site given by $\delta_{n_j, n'_j}$ is relaxed to represent some arbitrary overlap with the local Fock states on that site. This change is equivalent to defining the support configurations by product states with an arbitrary superposition over each local Fock space.
The kernel which compares these to the test configuration $\vec{n}$ now picks out its overlap with these local states in the superposition.
Following this through as shown in Ref.~\cite{boothQuantumGaussianProcess2021}, the wavefunction amplitudes then simply take the form
\begin{equation}\label{eq:qgps}
  \Psi_{\mathrm{qGPS}}(\vec{n}) = \exp \left( \sum_{n'=1}^M \prod_{i=1}^{L} \epsilon^{(n_i)}_{i, n'} \right),
\end{equation}
where the weights and definition of the support product states are subsumed into a single tensor of parameters, $\epsilon^{(n_i)}_{i, n'}$, shown in Eq.~\ref{eq:qgps}\footnote{The superscript index $n_i$ is a physical index (i.e., the local occupancy of the test configuration), the subscript index $i$ identifies the sites/modes, and $n'$ is an auxiliary index merely enumerating the support configurations or `latent space' of the model.}.
For a Fermionic system, where the local occupancy has a dimension of four, the ansatz is therefore fully parametrized by $4 \times L \times M$ continuous variational parameters, with $M$ playing the role of the systematically improvable parameter controlling the complexity of the state.
Having specified a set of parameters $\epsilon^{(n_i)}_{i, n'}$, the cost of evaluating a single configurational amplitude, $\vec{n}$, therefore scales as $\mathcal{O}(M L)$ (without any symmetrization of the model).

These more flexible support states defining the kernel function extend the variational freedom of the model, with the previous `classical' GPS representing a strict subset of the states that the `quantum' GPS span for the same value of $M$. We will denote this new form by the achronym `qGPS' where we need to distinguish it from its previous form supported by an explicit data set of simple `classical' particle configurations. %
As a further advantage of this new `quantum' form, the state is fully defined in terms of continuous parameters, allowing standard (variational) optimization techniques to be used to fully `learn' the model and smoothly transform between support configurations, avoiding the need for the coupled discrete-continuous optimization which necessitated the RVM for the optimization of the data set of support configurations.

Importantly, the qGPS representation is still applicable for a supervised learning of quantum states with Bayesian techniques, given a training set of configurational basis states and associated wavefunction amplitudes, ensuring that this perspective and valuable insight is not lost. To `learn' a qGPS model in this setup, Ref.~\cite{boothQuantumGaussianProcess2021} discusses a modification of the Bayesian training protocol which leverages the multi-linear character of the log-amplitudes of the model.
This iteratively updates the qGPS parameters one site at a time, sweeping through the physical space in a fashion conceptually related to matrix product state (MPS) optimization techniques such as DMRG~\cite{Schollwoeck2011} and TEBD~\cite{paeckelTimeevolutionMethodsMatrixproduct2019}.
From the perspective of tensor decompositions, this corresponds to an alternating least squares (ALS) approach~\cite{faberRecentDevelopmentsCANDECOMP2003, koldaTensorDecompositionsApplications2009, minsterCPDecompositionTensors2021} applied in the log wavefunction space, additionally incorporating Bayesian principles to compress the data into a compact model.
The additional Bayesian components allow for a systematic regularization of the fit, which is particularly helpful when only having access to limited training data and improving generalization of the model.

We observe the multilinear form for the log wavefunction amplitudes as
\begin{equation}
  \phi(\vec{n}) = \log \left( \Psi(\vec{n}) \right) = \sum_{n'=1}^M \prod_{i=1}^L \epsilon^{(n_i)}_{i, n'}.
\end{equation}
Due to the multilinearity, it is possible to extract one parameter per support index and local occupancy index as linear prefactors in a weighted sum of features defined by the other parameters.
In this construction, the log wavefunction amplitude is re-expressed as
\begin{equation}
  \phi(\vec{n}) = \sum_{n'=1}^M \sum_{l \in \{\cdot, \uparrow, \downarrow, \downarrow \uparrow\}} \epsilon^{l}_{I, n'} \, \delta_{n_{I}, l} \prod_{i \neq I} \epsilon^{(n_i)}_{i, n'},
\end{equation}
where the index $l$ labels the different local occupancies, and $I$ is a specific `central' index chosen as the reference site, whose parameters are updated.
With the identification of weights and features, the equation above can be written more compactly as a linear combination of features according to
\begin{equation}
  \phi(\vec{n}) = \sum_{i=1}^{M \times 4} w_i \, \chi_i(\vec{n}).
  \label{eq:linear_model_reference_site}
\end{equation}
The weights are given by the parameters associated with the reference site, $w_i = \epsilon^{l}_{I, n'}$, and the other parameters define the $4 \times M$ features $\chi_i(\vec{n}) = \delta_{n_{I}, l} \prod_{j \neq I} \epsilon^{(n_j)}_{j, n'}$  (where in both cases $i$ is a compound index of $n'$ and $l$).
With the reformulation of the qGPS model according to Eq.~\ref{eq:linear_model_reference_site}, the Bayesian regression techniques are directly applicable to obtain the weights $w_i$ in a well-defined, statistically meaningful approach from given wavefunction data.
The local regression can be iterated by repeatedly sweeping the choice of the reference site across the different modes of the system, updating $M$ parameters for each local occupancy value at a time.

Inspired by Ref.~\cite{Westerhout2019}, we consider the application of this Bayesian sweeping protocol for effective supervised learning from a limited data set~\cite{boothQuantumGaussianProcess2021}.
Having emerged as a common testbed, we consider an application to the rich phase diagram of the $J_1$-$J_2$ spin-1/2 model~\cite{nomuraDiractypeNodalSpin2021}, with frustrated magnetic orders notoriously hard to capture.
We trained the qGPS model using this Bayesian framework on a data set of a few randomly selected configurations and amplitudes from an exact ground state, before evaluating the quality of the state in terms of the overlap with the target state, where a high overlap can only be achieved by generalizing well beyond the training data.
The practical application of the sweeping scheme allows for a probabilistic interpretation of the result at each step, where the marginal likelihood can be used as an indicator of the quality of the fit, also taking into account the level of regularization to avoid overfitting of the data.
This allows the optimization of the prior and the noise level, which are updated during the sweeping to maximize the marginal likelihood in a \emph{type-II maximum likelihood} scheme~\cite{Tipping2003a}.

\begin{figure}
  \centering
  \includegraphics[width=\textwidth]{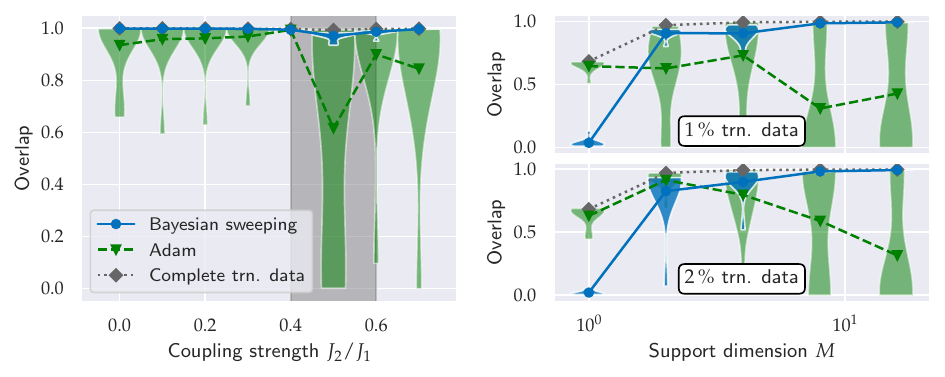}
  \caption{Overlap between a qGPS and the exact ground state, where the qGPS is trained from a small randomly selected configurational data set from this target ground state, for a $6 \times 4$ site $J_1$-$J_2$ square lattice. The results include a direct numerical minimization of the squared error with the Adam optimizer (green), and the application of the Bayesian sweeping (blue). Data shown includes the mean overlap (markers), and spread (violin plots), across ten different random realizations. The grey data points indicate the model expressiveness as obtained from training the model on the complete data set of all configurations and associated wavefunction amplitudes. Data in the left panel uses a model with a fixed support dimension of $M=5$ trained on $\approx 1 \%$ of data points from the full Hilbert space ( $\approx 2.7 \times 10^6$ basis configurations).
    The right panels show the results at fixed $J_2/J_1=0.5$ against the support dimension of the model for different training set sizes.
  Figure (adjusted) taken from Ref.~\cite{boothQuantumGaussianProcess2021}.}
  \label{fig:J1J2_sweeping_results}
\end{figure}

Figure~\ref{fig:J1J2_sweeping_results} shows the overlap evaluated between a learned qGPS wavefunction and the target state obtained for different parameter regimes of the $J_1$-$J_2$ model on a two-dimensional square lattice of $6 \times 4$ sites.
The plots visualize the distribution of the overlap for ten repetitions of the experiment at each parameter point to be able to extract the statistical fluctuations for different random realizations of random training data choices drawn from the target wavefunction.
The outcomes are presented for the principled Bayesian fitting approach (blue circles), as well as a squared error minimization with a stochastic gradient descent approach to optimize the fit (Adam)~\cite{kingmaAdamMethodStochastic2017} (green triangles).

When considering a qGPS with a fixed support dimension of $M=5$ (left panel), which was trained using random selections of only $1 \%$ of the full configurational space data, a significant spread of the results across different random seeds can be observed for the direct least squares minimization of the parameters.
On the other hand, the Bayesian sweeping approach gives a much more consistent fidelity to the target state, reaching the maximum possible overlap between the compressed qGPS representation and the target state in most cases.
Only within the frustrated regime shown in grey ($ 0.4 \lessapprox J_2/J_1 \lessapprox 0.6$), a slight increase in the spread across random realizations and a deviation from the model expressivity limit becomes apparent, pointing to the increase in complexity when capturing the structure of the wavefunction in frustrated regimes, as it was also observed in Ref.~\cite{Westerhout2019}.

This greater reliability of the Bayesian sweeping approach compared to more naive fitting strategies manifests in particular for larger support dimensions and more expressive models which are prone to overfitting with traditional approaches.
The right panel of Fig.~\ref{fig:J1J2_sweeping_results} reports the overlap in the strongly frustrated parameter regime at $J_2/J_1 = 0.5$ as the support dimension is increased from $M=1$ to $M=16$.
While a model with a small support dimension will typically not be sufficiently expressive to describe the target state well enough, appropriate regularization is crucial to avoid overfitting of the training data as the support dimension is increased towards full expressivity.
The Bayesian sweeping approach automatically incorporates appropriate regularization based on the data presented, resulting in a much improved reliability of generalizing beyond the training data as compared to alternative approaches, demonstrating the exceptional inductive bias of the model. This places the Bayesian GPS approach in a strong position for applications to quantum state tomography and beyond~\cite{koutnyNeuralnetworkQuantumState2022,Torlai2018a,Neugebauer2020,schmaleScalableQuantumState2021,ma2023attentionbased}.

\section{Ab initio quantum chemistry with machine learning-inspired quantum states}
\label{sec:3}
\subsection{Variational Monte Carlo}
The supervised learning task presented above highlights two main characteristics of the GPS framework:
\begin{enumerate}
    \item By tuning the support dimension, the model can be systematically improved to exactness, in principle able to represent any target quantum state.
    \item By deriving the model, kernel and supervised learning protocol from rigorous Bayesian principles and physically-inspired correlation characteristics, we can efficiently optimize a compact GPS wavefunction ansatz from a given configurational data set with good generalization properties, interpretable features and favourable inductive bias.
\end{enumerate}
However, to exploit the model characteristics for the determination of {\em unknown} quantum states relying on the variational principle, we can also consider the GPS as an ansatz within the context of variational Monte Carlo (VMC)~\cite{Becca2017}.
This relies on a sampling of states from the computational basis for a stochastic estimation of expectation values according to
\begin{equation}
\langle \hat{A} \rangle = \frac{\langle \Psi | \hat{A} | \Psi \rangle}{\langle \Psi | \Psi \rangle} = \sum_{\vec{n}} \frac{|\langle \Psi | \vec{n} \rangle|^2}{\langle \Psi | \Psi \rangle} \frac{\langle \vec{n} |\hat{A} | \Psi \rangle}{\langle \vec{n} | \Psi \rangle} = \sum_{\vec{n}} p_{\vec{n}} \, A_{loc}(\vec{n}).
\end{equation}
Here, the expectation value of an operator $\hat{A}$ is reformulated as the expectation value of configurational-local quantities, $A_{loc}(\vec{n}) = \frac{\langle \vec{n} |\hat{A} | \Psi \rangle}{\langle \vec{n} | \Psi \rangle}$, with respect to the probability distribution $p_{\vec{n}} = \frac{|\langle \Psi | \vec{n} \rangle|^2}{\langle \Psi | \Psi \rangle}$.
By sampling configurations from the Hilbert space according to the probability distribution $p_{\vec{n}}$, e.g. via application of the Metropolis-Hastings algorithm, and evaluating the mean of the local values over the sampled set, we obtain the estimate
\begin{equation}
\langle \hat{A} \rangle \approx \frac{1}{N_s} \sum_{\vec{n}_s} A_{loc}(\vec{n}_s).
\end{equation}
Crucially, the sum does not run over the full Hilbert space basis but a (typically) significantly smaller set of $N_s$ sampled configurations, and which should not need to scale with the Hilbert space size as dictated by the central limit theorem.
If the operator $\hat{A}$ is sparse in the chosen basis, i.e., each row in its matrix representation only has polynomially many non-zero entries, this average can be evaluated efficiently.
To find an approximate representation of a system's ground state (and the associated energy), we can then use the stochastic estimation of the energy expectation value (and its gradient) for a variational optimization of the parametrization by minimization of the variational energy.
An important property justifying this stochastic approximation of the energy is the zero variance principle.
This states that the variance over the local energies vanishes if the state corresponds to an eigenstate of the system.
Based on this, it can be expected that both the random and systematic errors of the expectation values should decreases as the trial state becomes a better representation of the target ground state. Further details of the VMC procedure and optimization techniques can be found in Ref.~\cite{Becca2017}.

\subsection{Gaussian Process States within Variational Monte Carlo}

Within the context of VMC, the optimization of unknown target states with the GPS model follows in similar footsteps as for Neural Quantum States (NQS), which have leveraged artificial neural network models as ansatz for the wavefunction~\cite{Carleo2017}.
Both descriptions rely on a systematically improvable function approximator to model the wavefunction amplitudes across the computational basis, which depend on a set of variational parameters that are optimized to approximate a certain target state.
A hyperparameter of the GPS model (in this case $M$) can be used to control the overall expressivity of the ansatz, which is linearly related to the number of (in general complex-valued) parameters as shown in Eq.~\ref{eq:qgps}.
This contrasts with NQS architectures, for which changing the expressibility typically entails changing the width and/or depth of the network architecture, but can also cover a large design space that also includes choices of the activation functions, pooling operations and connectivity between layers, all of which can change the span of accessible states in different ways.
Owing to its roots in kernel models, the GPS instead follows a particularly simple form derived from explicit physical principles in its design, and its variational span is specified by a single hyperparameter, $M$, controlling the flexibility of the model.

\begin{figure}
\centering
\includegraphics[width=1.0\textwidth]{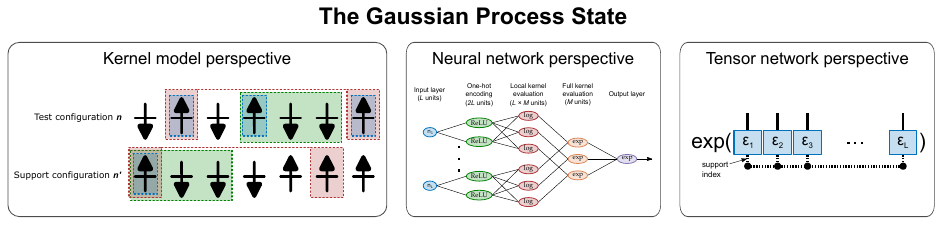}
\caption{Different perspectives can motivate the functional form of the GPS state, and be used to rationalize the states it can efficiently model, and motivate further developments in different domains. This includes its original genesis as a Bayesian kernel model, its recasting into a deep neural network architecture, and as the exponential of a matrix product state. In addition, constructive derivations from Jastrow and correlator product states are possible (see Ch.~\ref{sec:Bayesian_wavefunction})~\cite{boothQuantumGaussianProcess2021,rath_thesis}.}
\label{fig:GPS_perspectives}
\end{figure}

Mathematically, the functional form of the qGPS in Eq.~\ref{eq:qgps} is also equivalent to a tensor decomposition of the full $D^L$ tensor of the log wavefunction amplitudes, in terms of a sum over $M$ (tensor) products of $L$ one-dimensional tensors (where the one-dimensional tensors are indexed by the physical index $n_j$ labeling the states of the local Hilbert space of dimension $D$, with $D=4$ for the considered Fermionic systems).
Such a decomposition is known by various names, often simply denoted as a CP decomposition~\cite{kiersStandardizedNotationTerminology2000, koldaTensorDecompositionsApplications2009}. Alternatively, the log-wavefunction amplitudes can be thought of as a linear combination of $M$ product states, which can also be connected to matrix product states of bond dimension $M$, where all the matrices are constrained to be diagonal.
Such a construction for the actual wavefunction amplitudes as a simple linear combination would in general not fulfill the product separability requirements, however, the exponentiation of the CP decomposition ensures that a product of correlation features is obtained -- making it possible to represent common wavefunction ansatzes such as CPS or Jastrow states with compact support dimensions, while also substantially increasing the entanglement that the state can describe with an infinite resummation of linear combinations of product states. Some of these multiple perspectives on the GPS are summarized in Fig.~\ref{fig:GPS_perspectives}, and allow for insights and ideas to be transferred between fields. These enable cross-fertilization of developments to improve the scope, accuracy and optimization of the state in different domains, as we begin to explore below.

Building upon the GPS (or exponential of a CP decomposition) as a `parent' model, these adaptations can be incorporated to guide the efficiency and inductive bias of the model in practical applications.
We briefly highlight some of these extensions below:
\begin{enumerate}
    \item The model can be explicitly symmetrized to enforce spin, charge, point group and/or translational symmetries of the system. This can either be achieved by a `kernel-symmetrization' (i.e. symmetrizing the log-amplitudes) akin to the introduction of convolutional filters in NQS models~\cite{Bortone2024impactofconditional}, or by projective symmetrization schemes of the wavefunction amplitudes themselves, which have shown to help for the description of intricate sign structures with both neural networks~\cite{nomuraHelpingRestrictedBoltzmann2020} and the GPS~\cite{boothQuantumGaussianProcess2021}.
    \item The GPS can also be recast into a form which explicitly enforces normalization of the state. This is inspired by the construction of autoregressive neural network models. Importantly, this allows for a direct sampling of configurations without Markov chain approaches, therefore helping to prevent issues related to long autocorrelation times and loss of ergodicity~\cite{Sharir_2020}. This is based on a general functional form
\begin{equation}
\Psi_{\text{AR}}(\vec{n}) = \prod_{i=1}^L\frac{\Psi_i(n_i|\vec{n}_{<i})}{\sqrt{\sum_{d=1}^D|\Psi_i(n_i=d|\vec{n}_{<i})|^2}},
\end{equation}
where $\Psi_i(n_i|\vec{n}_{<i})$ is a `conditional' wavefunction of site $i$ that depends parametrically on the sub-configuration $\vec{n}_{<i}$, and the index $i$ defines a particular one-dimensional ordering of the degrees of freedom. Ref.~\cite{Bortone2024impactofconditional} provides a general recipe for the construction of an autoregressive ansatz integrated with the GPS model, however also pointing to a reduction of the expressiveness due to the restrictions imposed by the local normalization of the conditional wavefunctions~\cite{Bortone2024impactofconditional}. Extensions of the GPS to other autoregressive models can be achieved analogously, including recurrent neural networks~\cite{HibatAllah2020} and transformers~\cite{chen2025convolutionaltransformerwavefunctions}, which could also be envisaged with the GPS as the fundamental processing unit.
    \item The GPS state can be used as a Jastrow factor, supplemented by a multiplicative antisymmetric single-particle or pairing state which can be also efficiently evaluated, such as a single Slater determinant, Pfaffian~\cite{Bajdich2006} or antisymmetrized geminal power~\cite{Becca2017}. The GPS can then be used to modulate the many-body amplitudes of the state, explicitly capturing the many-electron correlations beyond the mean-field picture. This can be an effective strategy for describing Fermionic wavefunctions with the GPS in practice, including for lattice models~\cite{Glielmo2020}, and \emph{ab initio} quantum chemical systems~\cite{rath2023framework}.
    \item The CP decomposition can be a building block to construct backflow correlations within a Slater determinant or Pfaffian. Reference~\cite{bortone2024simplefermionicbackflowstates} constructed backflow correlations as many-body functions built on the CP decomposition to model configuration-dependent orbitals within the pairing state or single-particle Slater determinant. This approach has previously been extensively used within NQS in first quantization in real space~\cite{gerardGoldstandardSolutionsSchr2022, Pfau2019,Hermann2019, Han2018, hermannAbinitioQuantumChemistry2022, liInitioCalculationReal2022} and has since also found applications in the Fock space NQS representations~\cite{Luo2019, PhysRevB.110.115137, liu2025efficientoptimizationneuralnetwork}. Ref.~\cite{bortone2024simplefermionicbackflowstates} shows how the emerging inductive bias of the explicitly anti-symmetrized backflow construction utilizing the CP decomposition as ansatz for the configuration-dependent orbitals can help to avoid limitations for learning Fermionic states for lattice models and \emph{ab initio} systems. However, due to an increase in the computational cost of the model evaluation, the application has so far remained limited to small system sizes.
\end{enumerate}

We provide an overview of the different extensions to the GPS ansatz, their functional form, and the associated computational cost in table~\ref{tab:ansatzes_overview}.

\begin{table}
    \centering
    \begin{tabularx}{\textwidth}{ >{\centering\arraybackslash\hsize=0.16\hsize}X | >{\centering\arraybackslash\hsize=0.255\hsize}X | >{\centering\arraybackslash\hsize=0.35\hsize}X | >{\centering\arraybackslash\hsize=0.19\hsize}X}
         Ansatz &  Functional form & Model dependencies & Complexity \\
         \hline
         \hline
         `Classical' GPS~\cite{Glielmo2020} & \begin{gather*} \Psi(\vec{n}) = e^{\sum_{\{\vec{n}'\}} w_{\vec{n}'} \, k(\vec{n}, \vec{n}')}\end{gather*} & $\{\vec{n}'\}$: Support configurations;\newline $k(\vec{n}, \vec{n}')$: Kernel function;\newline $w_{\vec{n}'}$: Weights & \begin{gather*}\mathcal{O}[M L]\\(\mathcal{O}[M])\end{gather*} \\
         \hline
         `Quantum' GPS~\cite{boothQuantumGaussianProcess2021} & \begin{gather*} \Psi(\vec{n}) = e^{\sum_{n'=1}^M \prod_{i=1}^{L} \epsilon^{(n_i)}_{i, n'}}\end{gather*} & $\epsilon^{(n_i)}_{i, n'}$: Tensor of $D \times L \times M$ variational parameters & \begin{gather*}\mathcal{O}[M L]\\(\mathcal{O}[M])\end{gather*} \\
         \hline
         Kernel-symmetrized GPS~\cite{Glielmo2020, boothQuantumGaussianProcess2021} & \begin{gather*} \Psi(\vec{n}) = e^{\sum_{\mathcal{S}} \phi(\mathcal{S}[\vec{n}])}\end{gather*} & $\{\mathcal{S}\}$: $N_\mathcal{S}$ symmetry operations; \newline $\phi(\mathcal{S}[\vec{n}])$: Kernel model evaluated for transformed configuration $\mathcal{S}[\vec{n}]$ &
         \begin{gather*}\mathcal{O}[N_\mathcal{S} M L]\\(\mathcal{O}[N_\mathcal{S} M])\end{gather*} \\
         \hline
         Symmetry-projected GPS~\cite{boothQuantumGaussianProcess2021,nomuraHelpingRestrictedBoltzmann2020} & \begin{gather*} \Psi(\vec{n}) = \sum_{\mathcal{S}} X_\mathcal{S} \, e^{ \phi(\mathcal{S}[\vec{n}])}\end{gather*} & $\{\mathcal{S}\}$, $\phi(\mathcal{S}[\vec{n}])$: As above; \newline $X_\mathcal{S}$: Optional character prefactor for quantum number projection & \begin{gather*}\mathcal{O}[N_\mathcal{S} M L]\\(\mathcal{O}[N_\mathcal{S} M])\end{gather*} \\
         \hline
         Autoregressive GPS~\cite{Bortone2024impactofconditional} & \begin{gather*} \Psi(\vec{n}) = \\ \prod_{i=1}^L\frac{\Psi_i(n_i|\vec{n}_{<i})}{\sqrt{\sum_{d}|\Psi_i(d|\vec{n}_{<i})|^2}}\end{gather*} & $\Psi_i(n_i|\vec{n}_{<i})$: `Conditional' GPS for site~$i$; \newline $\vec{n}_{<i}$: Sub-configuration on sites preceding site~$i$ & \begin{gather*}\mathcal{O}[M L^2]\\(\mathcal{O}[ M L])\end{gather*}
         \\
         \hline
         Slater-Jastrow GPS~\cite{Glielmo2020, rath2023framework} & \begin{gather*} \Psi(\vec{n}) = \\ \Psi_{\mathrm{GPS}}(\vec{n}) \times \Psi_{\mathrm{MF}}(\vec{n})\end{gather*} & $\Psi_{\mathrm{GPS}}(\vec{n})$: GPS ansatz (quantum or classical); $\Psi_{\mathrm{MF}}$: Mean field ansatz (e.g. Slater determinant) & \begin{gather*}\mathcal{O}[M L + N^3]\\(\mathcal{O}[M + N])\end{gather*}\\
         \hline
         Backflow CPD~\cite{bortone2024simplefermionicbackflowstates} & \begin{gather*} \Psi(\vec{n}) = \\ \mathcal{A}[\phi_{\mu_11;\vec{n}} \ldots \phi_{\mu_NN;\vec{n}}]\end{gather*} & $\phi_{\mu_ii;\vec{n}}$: $i$-th backflow orbital parametrized by CP decomp.; \newline $\mathcal{A}$: Anti-symmetrization function & \begin{gather*}\mathcal{O}[N^2 M L + N^3] \\ (\mathcal{O}[N^2 M + N^3]) \end{gather*} \\
    \end{tabularx}
    \caption{Overview of wavefunction ansatzes derived from the GPS model. Evaluation complexity is specified by the scaling to evaluate a single amplitude with system size $L$, support dimension $M$, local Hilbert space dimension $D$, and electron number $N$. Scaling in parenthesis specifies the cost to apply a low-rank update to the amplitude, e.g. in the evaluation of local energies~\cite{boothQuantumGaussianProcess2021, rath2023framework,Becca2017}.}
    \label{tab:ansatzes_overview}
\end{table}

\subsection{Practical applications for ab initio quantum chemistry in second quantization}
\label{sec:quantchem_challenge}

We now return to the question of how GPS representations can be utilized for the study of \textit{ab initio} quantum chemistry in second quantization, outlined in section~\ref{sec:quantum_chemistry} and the general Hamiltonian definition of Eq.~\ref{eq:secondquantH}. We will highlight the choice of representation of the single-particle orbitals, and the effect this has on the ability to describe electronic ground states with the GPS model (and by extension NQS models~\cite{rath2023framework,Choo2019a, barrettAutoregressiveNeuralnetworkWavefunctions2021, zhaoScalableNeuralQuantum2022, https://doi.org/10.48550/arxiv.2301.03755, Yang2020, https://doi.org/10.48550/arxiv.2302.11588}).
We will specifically consider application to hydrogen systems with long-range interactions as a paradigmatic benchmarking testbed~\cite{stellaStrongElectronicCorrelation2011, sinitskiyStrongCorrelationHydrogen2010, mottaSolutionManyelectronProblem2017}.

\subsubsection{Basis choice}

The definition of the Hamiltonian in Eq.~\ref{eq:secondquantH}, relies on a specific choice of orthonormal basis functions that the indices label.
The representational power of a given GPS or NQS model is not invariant to this choice of basis, and the ability to represent the target state accurately can depend sensitively on this choice. In general, we can start from an orbital transformation from the underlying (non-orthogonal) atomic orbital basis, $\zeta_{j}(\mathbf{r})$, to a canonical or molecular orbital basis, $\Phi_{i}(\mathbf{r})$, as
\begin{equation}
\Phi_{i}(\mathbf{r}) = \sum_{\alpha=1}^L c_{\alpha i} \, \zeta_{\alpha}(\mathbf{r}).
\label{eq:orbital_linear_combination}
\end{equation}
where the coefficients of this molecular orbital expansion satisfy the canonical relation $\mathbf{c}^\dagger \mathbf{S} \mathbf{c}$, where $\mathbf{S}$ is the overlap matrix between the atomic orbital basis set. This matrix $\mathbf{c}$ can be obtained via the diagonalization of some effective one-electron Hamiltonian, such as that found in self-consistent-field calculations of Hartree--Fock (HF) or density functional theories.

However, this set of molecular orbitals is not unique, and for one valid set of molecular orbitals $\Phi_i$, another set of orbitals $\tilde{\Phi}_j$ can easily be obtained by applying a unitary transformation, characterized through a $L \times L$ matrix $U$ to give
\begin{equation}
\tilde{\Phi}_i = \sum_{j=1}^L U_{ij} \, \Phi_j,
\end{equation}
which still preserves the canonical orthogonality condition of the orbitals.
A unitary transformation of the molecular orbitals changes the structure of the many-body wavefunction amplitudes and will therefore influence the quality of the ground state approximation that can be achieved practically.
While it is generally not possible to identify one `most suitable' basis, we consider different heuristics in the construction of the orbitals, and distinguish their effect on the ability to represent electronic wavefunctions with machine learned models.

Taking orbitals from a mean-field (e.g. Hartree--Fock) level of theory results in a wavefunction which has the corresponding mean-field state described by a single many-body configuration, $\vec{n}_\mathrm{HF}$, corresponding to occupation of the lowest-lying HF orbitals when ordered energetically.
Assuming that the wavefunction ansatz can describe a single-configuration state, this choice of orbitals guarantees that the mean-field solution can easily be recovered as a ground state approximation within the machine learned ansatz.
For the GPS, for example, a support dimension of $M=1$ is sufficient to filter out the HF configuration and obtain a state with vanishing amplitudes on the remainder of the Hilbert space~\cite{rath2023framework}. As a result, when using a canonical orbital representation, the target state will have a particularly peaked structure around the HF configuration for weakly to moderately correlated systems, which may lead to limitations in a faithful sampling of the configurational space and the ability to generalize the optimization of ansatzes with few Monte Carlo samples~\cite{Choo2019a}.

Motivated by the success of machine learned ansatzes for lattice models, one might alternatively consider a construction of the orbitals based on a notion of locality in real space.
With this aim, we can construct an expansion of atomic orbitals such that the final orbitals are as localized as possible.
Different approaches can be applied to achieve this goal in practice.
These are typically either based on closed form expressions~\cite{lowdinNonOrthogonalityProblem1950,reedNaturalPopulationAnalysis1985,aquilanteFastNoniterativeOrbital2006}, or on a numerical minimization of a metric quantifying the locality of the molecular orbitals~\cite{fosterCanonicalConfigurationalInteraction1960,edmistonLocalizedAtomicMolecular1963,pipekFastIntrinsicLocalization1989}.
An example of localized orbitals is the basis of `Boys'-localized orbitals~\cite{kleierLocalizedMolecularOrbitals1974}, which we consider in the following.
The unitary rotation matrix $U$, transforming an initial set of molecular orbitals to the localized ones, is determined by numerical minimization of the measure
\begin{equation}
\mathcal{L}(U) = \sum_{i=1}^L \left | \int d \mathbf{r} \, \tilde{\Phi}^\ast_i(\mathbf{r}) \, \mathbf{r} \, \tilde{\Phi}_i(\mathbf{r}) \right |^2.
\end{equation}
This is equivalent to the construction of Wannier functions common in the solid state. For a localized representation, the overall state will generally have less weight concentrated in few configurations and have a broader distribution across the computational basis.
Furthermore, as the molecular orbitals become more localized, the amount that they overlap will decrease for all pairs of orbitals.
As a consequence, the two-electron interaction term of the Hamiltonian becomes sparser, asymptotically leaving only $\mathcal{O}(L^2)$ non-vanishing terms in the two-electron integral tensor $v_{ijkl}$.
The use of localized orbitals therefore also enables potential routes to push the practical applicability of second quantized wavefunction parametrizations to larger systems where the computational cost of the local energy evaluations is in-line with descriptions in real-space~\cite{weiReducedScalingHilbert2018, foulkesQuantumMonteCarlo2001}.

\begin{figure}
\centering
\includegraphics[width=0.8\textwidth]{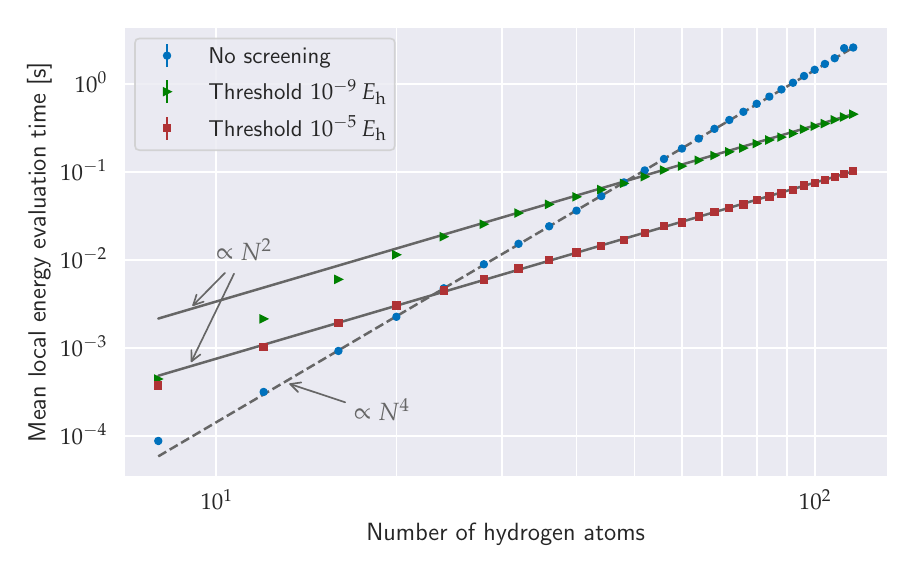}
\caption{Mean local energy evaluation time as a function of the number of atoms for a linear chain of hydrogen atoms with fixed inter-atomic separation of $1.8 \, a_0$ (minimal basis). Blue points indicate no screening of the Hamiltonian matrix elements, while red and green points screen the Hamiltonian elements with a threshold of $10^{-5} $ $E_\mathrm{h}$ and $10^{-9}$ $E_\mathrm{h}$ respectively. Figure (adjusted) taken from Ref.~\cite{rath2023framework}.}
\label{fig:H_chain_scaling analysis}
\end{figure}

In Fig.~\ref{fig:H_chain_scaling analysis}, we demonstrate this asymptotic $\mathcal{O}(N^2)$ scaling for the local energy in a local basis~\cite{rath2023framework}.
It shows the mean computational time required for the evaluation of the local energy, involving the summation over all connected basis states for a sampled configuration.
We consider the dependence on the number of atoms in a chain of hydrogen atoms with a fixed inter-atomic spacing, where the mean time is evaluated using an ansatz which can be evaluated in $\mathcal{O}[1]$ time.
As expected, an evaluation of the full quartic number of Hamiltonian terms results in a scaling of $\mathcal{O}[N^4]$, which is reduced to  $\mathcal{O}[N^2]$ by pruning terms with magnitude falling below a threshold and utilizing a sparse representation of the Hamiltonian.
While this practically comes with a small overhead, the reduced scaling gives a computational advantage for system sizes of $\approx 25 - 50$ atoms, depending on the truncation threshold.

\subsubsection{Benchmarking applications for strongly correlated hydrogen materials}

\begin{figure}
\centering
\includegraphics[width=\textwidth]{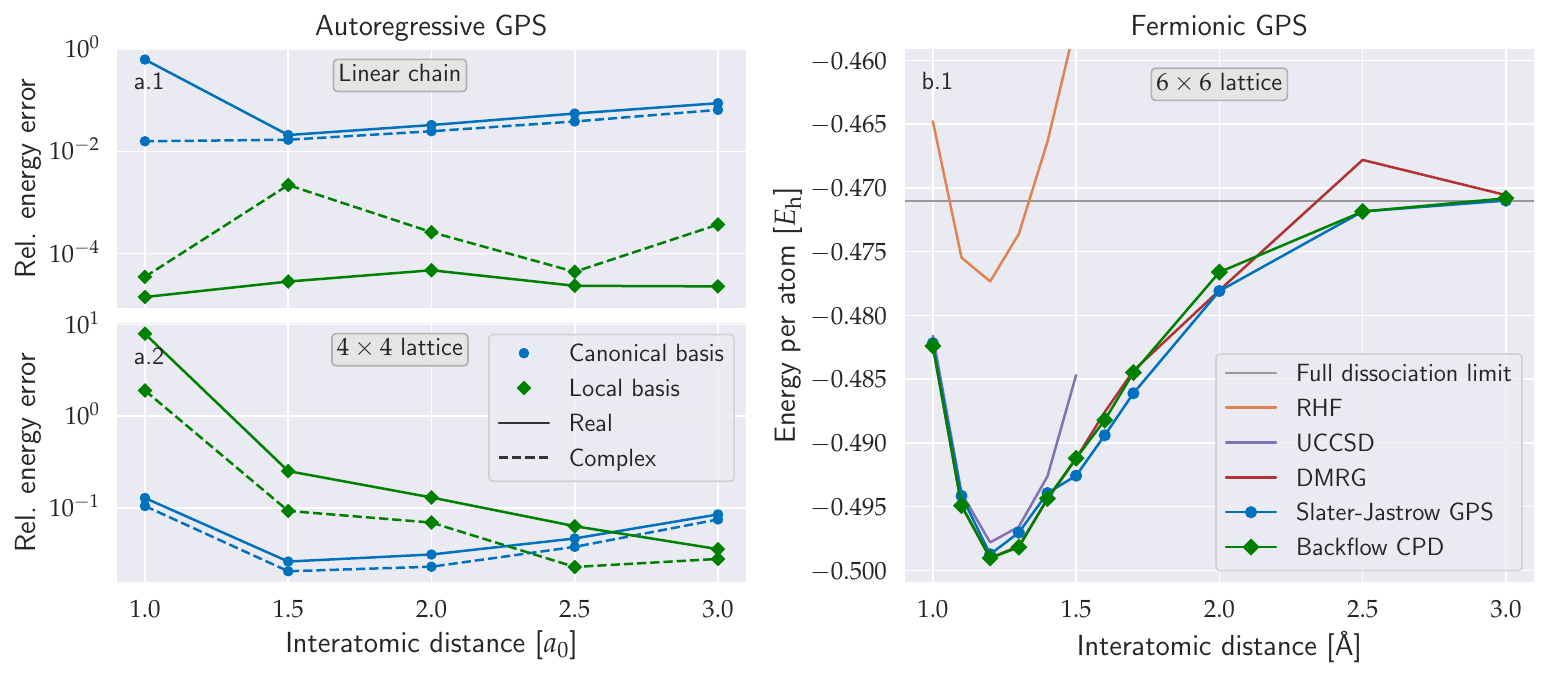}
\caption{Ground state potential energy surface with different variants of the GPS model for one- and two-dimensional symmetrically stretched arrays of hydrogen atoms (minimal basis set). Left panel: Relative energy error as a function of the interatomic distance obtained for small systems with the autoregressive GPS~\cite{Bortone2024impactofconditional} with real (solid) and complex (dashed parameters) in a canonical basis of Hartree-Fock orbitals (blue) and a local basis of Boys orbitals (green), from Ref.~\cite{Bortone2024impactofconditional}. Right panel: Potential energy surface for a GPS model supplemented with a single Slater determinant~\cite{rath2023framework} (blue) and a backflow model based on the CP decomposition~\cite{bortone2024simplefermionicbackflowstates} (green), as well as additional reference results, for a two-dimensional array of $6 \times 6$ hydrogen atoms from Ref.~\cite{bortone2024simplefermionicbackflowstates}.}
\label{fig:H_materials_comparison}
\end{figure}

With the set up of the {\em ab initio} electronic structure problem in second quantized Fock space, the different GPS variants can approximate ground state properties within the VMC framework.
The success of autoregressive GPS variants for the  description of small one- and two-dimensional arrays of hydrogen atoms is shown in the left panel of Fig.~\ref{fig:H_materials_comparison}, taken from Ref.~\cite{Bortone2024impactofconditional}.
For a one-dimensional array of 16 hydrogen atoms described in a minimal local basis, a relative error of less than $10^{-4}$ is achieved with an autoregressive GPS ($M=16$) across the simultaneous symmetric stretching of all bonds.
This significantly outperforms the result obtained for the description of the state in the canonical basis of Hartree-Fock orbitals, pointing to an increased complexity with representing and learning the wavefunction in this basis. Similar results can also be achieved for the non-autoregressive GPS model~\cite{boothQuantumGaussianProcess2021}.

However, this accuracy largely derives from the quasi-one-dimensional nature of this system, which enables the wavefunction to be described largely free from an intricate sign structure in a local basis representation. In the limit of a fully 1D Hubbard model with local interactions, the ground state can be represented free from any sign structure at all.
Within the autoregressive GPS, this quasi-1D form manifests in the autoregressive GPS being able to be described largely with real-valued parameters, which are significantly easier to optimize and represent.
Strong modulation of the sign structure across the Hilbert space, as required in higher-dimensional systems, requires complex parameters and introduces additional challenges in the faithful optimization of the state, marked by larger errors in the local basis compared to the sign-structure-free counterpart.

Also shown in Fig.~\ref{fig:H_materials_comparison} for the 1D linear chain are results from the autoregressive GPS model for a canonical (HF) orbital representation. Despite still being a 1D system, the results are materially different, with the non-local basis choice introducing an intricate sign structure in the modeled target wavefunction.
While the signed ansatz with complex parameters can formally model this structure, the relative energy error is significantly larger across the full range of stretched conformations. Moving to the 2D $4 \times 4$ array of atoms (a.2), we find that neither of the representations can reach the same accuracy as in the 1D case, with non-trivial sign structures in both a local and canonical orbital representations. This trend has similarly emerged in other settings: The accurate representation of signed target states still represents one of the key challenges in the application of machine learned quantum states~\cite{szabó_2021, westerhoutUnveilingGroundState2022}.

Alternatively, modified ansatzes can incorporate the Fermionic character directly into a GPS-based ansatz.
While the second quantized formalism does not strictly necessitate an explicit anti-symmetrization of the ansatz, this has proven key to being able to capture the electronic structure of general quantum chemical systems accurately. These models build upon the ability to capture the mean-field and antisymmetric sign structure characteristics of the target state efficiently, without encoding them into the GPS correlations directly.
In the right panel of Fig.~\ref{fig:H_materials_comparison}, we highlight the performance of the GPS model supplemented with a Slater determinant reference state as well as the backflow model in which the orbitals are directly described as configuration-dependent many-body configurations for a two-dimensional system of $6 \times 6$ hydrogen atoms.
Being outside the scope of a numerically exact reference calculations (FCI), the panel b.1 shows the total potential energy of the system as the bonds are symmetrically expanded.
Across all geometries, the GPS model, modulating a correction to a reference mean-field state, achieves good agreement with the backflow model, both giving potential energies within a spread of less than $0.003 \, E_\mathrm{h}$.
The consistency and quality of the two models is also highlighted by the good agreement with density matrix renormalization group (DMRG) results~\cite{Block}, providing another variational estimate of the energy.
Whereas the DMRG solution fails to achieve an appropriately representative variational energy at $2.5 \,$\AA, the GPS-based models give a variational energy estimate in much better agreement with other parts of the dissociation curve.

\begin{figure}
\centering
\includegraphics[width=0.8\textwidth]{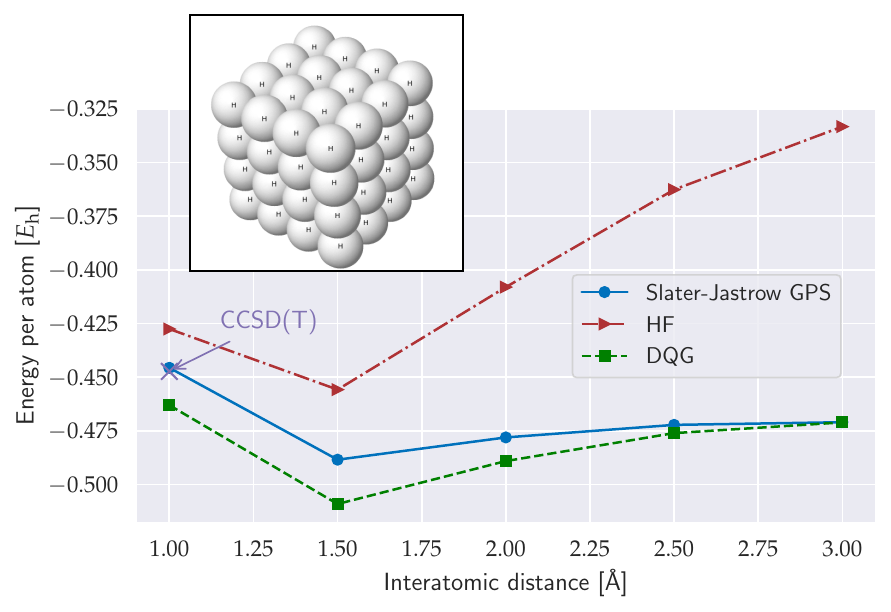}
\caption{Ground state potential energy from a GPS with Slater determinant reference state for a symmetrically stretched three-dimensional array of $4 \times 4 \times 4$ hydrogen atoms as a function of the inter-atomic separation in a minimal basis set (STO-6G). Comparison values include Hartree-Fock (red) and the anti-variational two-body reduced density matrix methods with approximate DQG representability constraints~\cite{sinitskiyStrongCorrelationHydrogen2010}, as well as a CCSD(T) reference value (purple). Figure (adjusted) taken from Ref.~\cite{rath2023framework}.}
\label{fig:H_cube_results}
\end{figure}

The necessity of the Fermionic structure to be encoded explicitly into the model has also been observed in the study of other molecular structures, such as a relaxed H$_2$O molecule~\cite{rath2023framework, bortone2024simplefermionicbackflowstates}.
While the backflow construction has shown to improve upon solutions obtained with a GPS augmented mean-field ansatz for systems displaying intricate Fermionic sign structures~\cite{bortone2024simplefermionicbackflowstates}, its additional computational complexity pose additional challenges for a scaling of the approach, which currently remain largely unaddressed.
Representing the largest application of a comparable ansatz to date, Ref.~\cite{rath2023framework} showcases an application of the GPS augmented Slater determinant ansatz for a three-dimensional system of $64$ hydrogen atoms ($128$ spin orbitals).
These results shown in Fig.~\ref{fig:H_cube_results} correspond to a variationally optimized qGPS, with a support dimension of $M=96$, augmented by a single Slater determinant reference state.
No methods exist to obtain accurate reference energies for this system across the range of symmetrically stretched conformations, but agreement with CCSD(T) at a compressed geometry (the only geometry considered where this `gold-standard' quantum chemistry approach appropriately converged) and lower-bound variational-2RDM methods at stretched geometries~\cite{sinitskiyStrongCorrelationHydrogen2010} provide confidence that this represents a new state-of-the-art result.

The applications of these hydrogen systems, which can tune both dimensionality and weak-to-strong correlation strengths, are a stern test for different variants of the GPS ansatz when applied to quantum chemistry with long-range interactions. The conclusions drawn are also likely extendable to other systematically-improvable machine learning ansatzes, and sketch out a direction for further work in accurate ab initio quantum chemical simulations with these states.
While applications in a second quantized formalism with VMC have shown that systematic improvability of the approximation can not always be guaranteed, the ability to capture the characteristics of strongly correlated electronic quantum states promises new potential for predicting the chemical behaviour on the quantum scale in regimes where other approaches fail. While first quantized NQS variants have often demonstrated the best accuracy for molecular systems, we believe that further methodological developments could place second quantized representations as a viable alternative, and are necessary to realize the full potential of these states.

\section{Quantum-inspired machine learning}

The GPS framework defined above introduces a family of specific ansatzes to model particular many-body quantum states.
While its derivation is focused on quantum many-body systems, the ansatz itself can also be seen as a more general functional form associating a scalar value to vectorial inputs.
It is therefore interesting to consider whether these `quantum-inspired' mappings can be an effective representations of input-output relationships in  other contexts, where features similar to correlations of interacting quantum states emerge within data.
This is somewhat analogous to tensor network states, which have recently found significant interest as machine learning models beyond the description of quantum states~\cite{stoudenmireSupervisedLearningQuantumInspired2017, hanUnsupervisedGenerativeModeling2018, bradleyModelingSequencesQuantum2019, glasserExpressivePowerTensornetwork2019, bhatiaMatrixProductState2019, chengSupervisedLearningProjected2020, dymarskyTensorNetworkLearn2021, convyMutualInformationScaling2021, liuTensorNetworksUnsupervised2021, luTensorNetworksEfficient2021, linTensorNetworkSupervised2021, barrattImprovementsGradientDescent2022, baiUnsupervisedRecognitionInformative2022, senguptaTensorNetworksMachine2022, vieijraGenerativeModelingProjected2022, strashkoGeneralizationOverfittingMatrix2022, sommerEntanglingSolidSolutions2022, zunkovicDeepTensorNetworks2022, helmsDynamicalPhaseBehavior2019, banulsUsingMatrixProduct2019, causerOptimalSamplingDynamical2022, jahromiVariationalTensorNeural2022, Stokes_2019, Liu_2019, https://doi.org/10.48550/arxiv.2212.14076, millerProbabilisticGraphicalModels2021, Otgonbaatar2023, Sun_2020, chen2023machine}. This reverses the direction of inspiration which found neural networks as function approximators in quantum many body systems. Similar to the application of tensor networks, we can ask the question of whether the GPS framework may be effective as a model in other machine learning contexts. This certainly seems plausible, since it has the features of the transformations in neural network approximators of a (multi-)linear transformation followed by a non-linearity (the exponential). Indeed, it has been previously shown how the GPS can be brought into the form of a neural network model with specific connectivity and activation functions~\cite{boothQuantumGaussianProcess2021}.

To begin these investigations, in this section we apply the Bayesian sweeping supervised learning algorithm discussed in Sec.~\ref{sec:qGPS} to a common image recognition task with the qGPS as a model~\cite{rath_thesis}.
The setup is inspired by Ref.~\cite{stoudenmireSupervisedLearningQuantumInspired2017} which demonstrated the use of an MPS model for this task.
Due to relation between the GPS model and CP decompositions, supervised learning approaches with the model are related to similar techniques leveraging the compression ability of this decomposition for different machine learning applications, commonly used in conjunction with artificial neural networks~\cite{lebedevSpeedingupConvolutionalNeural2015, caoTensorRegressionNetworks2018, jiFastCPcompressionLayer2022}.
This shows how the different perspectives brought together in the GPS model, namely Bayesian regression principles, fundamentals of many-body wavefunction modeling, and tensor decompositions, can provide universal tools for such supervised learning tasks.

We consider the classification of the digits from the MNIST data set~\cite{lecunGradientbasedLearningApplied1998}, comprising $28 \times 28$ greyscale pixels representing possible numbers.
This digit recognition task represents a simple supervised learning classification problem.
A set of training examples is used to train the model and associate images with one of the ten different digit classes.
Being a prototypical setup for a practically relevant classification task, learning from and testing methods on the MNIST data set has become a standard benchmarking for different methods and models, including ones inspired by tensor network representations~\cite{stoudenmireSupervisedLearningQuantumInspired2017, hanUnsupervisedGenerativeModeling2018, chengSupervisedLearningProjected2020, convyMutualInformationScaling2021, dymarskyTensorNetworkLearn2021, liuTensorNetworksUnsupervised2021, baiUnsupervisedRecognitionInformative2022, strashkoGeneralizationOverfittingMatrix2022, zunkovicDeepTensorNetworks2022, Liu_2019, Sun_2020}.

For the application of the GPS model, the MNIST dataset comprises 60,000 training images, and 10,000 further images to validate the accuracy.
By analogy to quantum many-body systems, each image can be flattened to represent a configurational vector $\vec{n}$ for which the element $n_i$ gives the greyscale value of the $i$-th pixel.
To apply the GPS, a `one vs. rest' approach~\cite[section 7.6]{scholkopfLearningKernelsSupport2001} is followed, where a separate model is introduced for each digit class.
Ten different GPS models, $\Psi^{(d)}_\mathrm{GPS}(\vec{n})$, are therefore trained for each of the ten digit classes $d = 1 \ldots 10$, with each model representing a GPS-style mapping from the input to an (unnormalized) probability determining whether the input is considered to be element of that class or not.

In applications of the GPS to quantum systems, the elements of the input vectors $\vec{n}$ took one out of $D$ values (with $D$ being the dimensionality of the local Hilbert space, which was $4$ in the case of electronic spin orbitals).
For the considered digit classification however, the vector elements are continuous greyscale values (in the case of the MNIST dataset, represented with a precision of eight bits).
The core element of the GPS is to construct the functional estimator as an (exponentiated) linear combination of $M$ support points described as product states.
To extend this to quasi-continuous local degrees of freedom (such as the greyscale value), different approaches are possible to parametrize the local values via a general function $f_{d, i, x'}(n_i)$, often called an `embedding' of the data.
This function will associate an amplitude to a local greyscale value $n_i$ for pixel $i$, support point $x'$, and digit class $d$.
While further investigations are required to assess the influence of different embeddings on the final results, here, a simple linear model is assumed for the local state.
This parametrizes the state in the fashion of a visible unit of a neural network as
\begin{equation}
f_{d, i, x'}(n_i) = \epsilon^{(0)}_{d, i, x'} + \epsilon^{(1)}_{d, i, x'} n_i,
\end{equation}
where $\epsilon^{(0)}_{d, i, x'}$ and $\epsilon^{(1)}_{d, i, x'}$ are the variational parameters associated with the ansatz for digit class $d$.
Alternative choices for the input encoding could, e.g., be obtained by discretizing the greyscale value, or by encoding the greyscale value as a local spin rotation, a construction used in Ref.~\cite{stoudenmireSupervisedLearningQuantumInspired2017}. Overall, this requires a two-dimensional local Hilbert space of parameters to encode the greyscale value, $n_i$.

Based on the chosen embedding, the functional model for the classification is defined as
\begin{equation}
\Psi^{(d)}_\mathrm{GPS}(\vec{n}) = \exp \left(  {\sum_\mathcal{S} \sum_{x'=1}^M \prod_{i=1}^L (\epsilon^{(0)}_{d, i, x'} + \epsilon^{(1)}_{d, i, x'} \mathcal{S}[n]_i)} \right),
\end{equation}
where $L$ is the total number of pixels.
This ansatz includes an additional (generally optional) sum over symmetry operations, $\mathcal{S}$, which can be included to symmetrize the model according to a 'kernel-symmetrization' approach.
In the discussed setup of image classification, we consider the symmetry operations given by all shifts of the image by up to two pixels in any direction, therefore giving a total of $25$ considered symmetry operations.
For the translational shifts of the image data, white pixels were added at the opposite side of pixels shifted across the boundary (of which the values are discarded).
Other symmetrization approaches could be considered in the future,
for example by full symmetrization of the model according to all translations with assumed periodic boundary conditions~\cite{luTensorNetworksEfficient2021}, or by incorporating other symmetry operations such as rotations of the image~\cite{byerlyNoRoutingNeeded2021, anEnsembleSimpleConvolutional2020}.

Since the (unnormalized) probabilities $\Psi^{(d)}_\mathrm{GPS}(\vec{n})$ are still exponentiated multilinear models, an application of an iterative Bayesian sweeping is readily applicable to learn the probability models from the available training data.
To directly apply the regression of the models, the models are fit on probability amplitudes, either vanishing if the training configuration is not associated with the class, or giving a value of one if they are.
Each of the ten different GPS models can then be fit with the Bayesian sweeping approach on the set of training images, $\{\vec{n}\}_\mathrm{tr}$, with a set of log training amplitudes $\{\omega^{(d)}\}$.
To facilitate the fitting in the log space of the probabilities, we set vanishing amplitudes to a small value, chosen as the (approximate) variance of the amplitude likelihood, $\tilde{\sigma}^2$, which is a hyperparameter updated by maximization of the marginal likelihood to denote the intrinsic uncertainty in the classification.

Having optimized the models with a Bayesian sweeping protocol (which may be further fine-tuned for a probabilistic description and uncertainty quantification of class labels~\cite{Rasmussen2006}), we can easily predict digit labels for other image inputs.
To classify an image $\mathbf{n}$ (potentially one not included in the training data), the probability amplitudes $\Psi^{(d)}_\mathrm{GPS}$ are evaluated for all classes, and a label is predicted according to the class for which the evaluated amplitude is the largest.

\begin{figure}
\centering
\includegraphics[width=\textwidth]{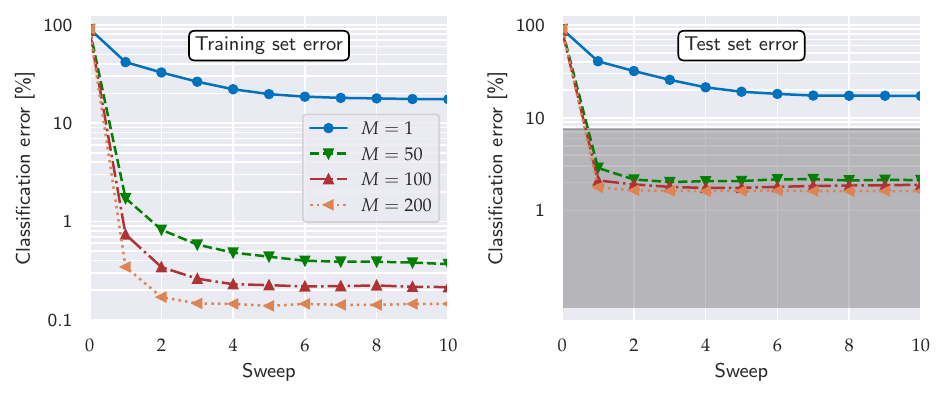}
\caption{Percentage of incorrectly classified images from the MNIST data set in relation to the number of sweeps applied to train GPS models with support dimensions $M=1$ (blue data points), $M=50$ (green data points), $M=100$ (red data points), and $M=200$ (orange data points). The left plot shows the classification error of the training inputs, the right plot shows the error for the classification of the test images. The shaded area in the right plot denotes the range between the overall lowest test error of $0.09 \%$ misclassified images achieved with the approach from Ref.~\cite{anEnsembleSimpleConvolutional2020}, and the highest test error rate ($7.53 \%$~\cite{shi2022personalized}) from the comparison of state-of-the-art approaches listed on \url{https://paperswithcode.com/sota/image-classification-on-mnist} (last accessed on 4/6/2023).}
\label{fig:MNIST_results}
\end{figure}

Figure~\ref{fig:MNIST_results} shows the percentage of misclassified images from the MNIST data set in relation to the number of sweeps across all pixels applied in the Bayesian training using different GPS support dimensions $M=1,50,100,200$.
The left plot reports the classification error obtained for the prediction of the labels from the training set.
It can be seen that, for all displayed support dimensions, the training set error decreases rapidly and convergence is observed after few sweeps.
Furthermore, the approached value shows systematic improvement with respect to increases in the support dimension of the models.
With a support dimension of $M=1$, slightly less than $20 \, \%$ of the images from the training set are not correctly classified after ten sweeps.
This error decreases to a value of $\approx 0.4 \, \%$ for the model with $M=50$, and a value of $\approx 0.1 \, \%$ for the model with $M=200$.

While the quick decrease to small errors on the training set indicates that the training data is fit appropriately, the key quantity of interest is how well the representations generalize outside the training data.
The right plot of the figure shows the achieved error across the $10,000$ unseen test images of the MNIST data set.
The percentage of incorrectly classified test images also overall shows a rapid decrease during the first few sweeps for all considered support dimensions $M$.
Whereas the test set error is approximately equal to the training set error for the simplest model with $M=1$, slight discrepancies of about $2 \%$ between training and test errors can be observed for the more expressive models, pointing to a small degree of overfitting of the models with $M=50,100,200$ to the training data, as it has also been discussed for other tensor network protocols~\cite{strashkoGeneralizationOverfittingMatrix2022}.
Nonetheless, the overall test accuracies approached still show marginal improvements for the considered increases of the support dimension, and test error rates of $\approx 1.6 - 1.7 \, \%$ are achieved with $M=200$ after three sweeps.
While the MNIST data represents a rather simple data set, the introduced scheme is relatively general and the presented results already provide a first indication of a general applicability of the method.
The classification accuracies achieved for the MNIST dataset do not reach the overall highest accuracies achieved with some ML approaches often especially fine-tuned for the task at hand and potentially including further augmentation of the training set~\cite{byerlyNoRoutingNeeded2021}.
For the MNIST dataset, test accuracies as small as $0.09 \%$ have been reported~\cite{anEnsembleSimpleConvolutional2020}.
Nonetheless, the accuracies obtained with the discussed Bayesian sweeping method are within the range of other state-of-the-art results.
For comparison, the performance benchmarks listed on the website \url{https://paperswithcode.com/sota/image-classification-on-mnist} (last accessed on 4/6/2023) include test set classification error rates of state-of-the-art methods introduced between 2013 and 2023 ranging from $0.13 \%$~\cite{byerlyNoRoutingNeeded2021} up to $7.53 \%$~\cite{shi2022personalized}.
The range of test accuracies achieved with different state-of-the-art methods is indicated in the right plot of fig.~\ref{fig:MNIST_results} where the region between the lowest test error rate of $0.09 \%$~\cite{anEnsembleSimpleConvolutional2020} and the test error rate of $7.53 \%$~\cite{shi2022personalized} is shaded.
Overall, this application of the GPS model for digit classification from scanned images provides a clear example of how the synergies between the fields of computational quantum science and machine learning may prove to underpin future advancements in both fields.

\section{Perspectives}

One of the central questions within the field of quantum chemistry is how quantum many-body wavefunctions can be modeled and efficiently optimized to allow for accurate solutions of the electronic structure problem.
Within this chapter, we have discussed a wavefunction ansatz derived from Bayesian machine learning principles, the Gaussian Process State, and its application to model the ground state of quantum many-body systems with intricate quantum correlations. Finally, we considered the scope for recasting these quantum ansatz back as machine learning models in more general contexts.

The success of the description is rooted in two fundamental modeling principles incorporated into the construction of the state.
Firstly, the state derives from strict product separability of correlation features, allowing for non-trivial compact representations of the state for larger system sizes.
Secondly, a careful design of a kernel function, describing a co-variance between function points in the Gaussian process regression picture, makes it possible to incorporate explicit physicality into the model, without imposing a priori restrictions of the overall state expressivity.

The GPS is in spirit very similar to the family of Neural Quantum States (NQS) relying on the representative power of artificial neural networks.
These functional forms employed as wavefunction ansatz generally do not allow for exact deterministic contraction of expectation values.
This leads to the requirement to apply stochastic sampling approaches to compute physical quantities within the framework of VMC.
Numerical approaches are thus inherently influenced by noise of the estimation procedures, which can become a practical hindrance in reliably uncovering physical behavior of interest.
Only if quantum ansatzes can be learned efficiently based on a limited and practically feasible numbers of samples, the VMC framework can offer a viable route to tackle the many-body problem for application in the most challenging application cases in quantum chemistry.
The observed difficulties are not necessarily specific to the GPS ansatz, but they match those observed in the practical application of NQS, where both families appear to suffer from similar limitations with being able to learn the representation faithfully from a limited set of configurational samples.

While it is generally hard to separate limitations in the representative power of the model and its numerical optimization, there is increasing evidence that the latter is at least part of the wider problem.
Due to its roots in rigorous Bayesian learning frameworks and being derived from physical principles, the GPS might offer wider solutions to the core optimization challenges within the field, though these have not yet seen general success in quantum chemistry. However, if this can be overcome, then advantages of working in second quantization can be found, including the ability to integrate with Fock space multi-scale resolution schemes and techniques of explicit correlation.
The presented GPS framework offers a promising new direction towards overcoming the foundational challenges towards the dream of a universal toolbox to accurately predict chemical behavior from first principles.

\begin{acknowledgement}
The authors gratefully acknowledge support for this project from the Air Force Office of Scientific Research under award number FA8655-22-1-7011. YR furthermore acknowledges support by the Department of Science, Innovation and Technology’s International Science Partnership Fund (ISPF).
\end{acknowledgement}
\ethics{Competing Interests}{
The authors have no conflicts of interest to declare that are relevant to the content of this chapter.}

\eject

\printbibliography
\end{document}